\documentclass{ws-mplb}
\usepackage{psfrag,cite}
\begin{document}

\markboth{Bal\'azs D\'ora, Kazumi Maki, Attila Virosztek}{Recent Advances in Unconventional Density Waves}

%
\catchline{}{}{}{}{}
%

\title{Recent Advances in Unconventional Density Waves}

\author{Bal\'azs D\'ora}

\address{The Abdus Salam ICTP, Strada Costiera 11, I-34014, Trieste, Italy\\
dora@ictp.trieste.it}

\author{Kazumi Maki}

\address{Department of Physics and Astronomy, University of Southern California, Los Angeles CA 90089-0484, USA}

\author{Attila Virosztek}

\address{Department of Physics, Budapest University of Technology and Economics, H-1521 Budapest, Hungary and 
Research Institute for Solid State Physics and Optics, P.O.Box 49, H-1525 Budapest, Hungary}

\maketitle

\begin{history}
\received{(received date)}
\revised{(revised date)}
\end{history}

\begin{abstract}
Unconventional density wave (UDW) has been speculated as a possible electronic ground state in excitonic 
insulator in 1968.
Recent surge of interest in UDW is partly due to the proposal that the pseudogap phase in high $T_c$ cuprate
superconductors is d-wave density wave (d-DW).

Here we review our recent works on UDW within the framework of mean field theory.
In particular we have shown that many properties of the low temperature phase (LTP) in
$\alpha$-(BEDT-TTF)$_2$MHg(SCN)$_4$ with M=K, Rb and Tl are well characterized in terms of unconventional
charge density wave (UCDW). In this identification the Landau quantization of the quasiparticle motion in a 
magnetic field
(the Nersesyan effect) plays the crucial role. Indeed the angular dependent magnetoresistance and the 
negative giant Nernst effect are 
two hallmarks of UDW.
\end{abstract}

\section{Introduction}

Until recently the electronic ground states in  crystalline solids are considered to belong to one of four canonical ground states in 
 quasi-one dimensional systems: s-wave superconductors, p-wave superconductors, (conventional) charge density wave and (conventional) 
spin density wave\cite{solyom,jerome,gruner}. Indeed many systems discovered since 1972 appeared to accommodate in this scheme: CDW in 
NbSe$_3$ and SDW in 
Bechgaard salts (TMTSF)$_2$PF$_6$\cite{ishiguro}. In all of these systems the quasiparticle spectrum has the energy gap $\Delta$ and 
the quasiparticle 
density decreases exponentially as $e^{-\Delta/T}$ as the temperature decreases to $T\ll \Delta$. Also the thermodynamics of these 
systems are practically the same as the one for s-wave superconductors as described by the theory of Bardeen, Cooper and 
Schrieffer\cite{BCS} (i.e. the BCS theory).

However since the discovery of heavy fermion superconductors, organic superconductors, high $T_c$ cuprate superconductors and 
Sr$_2$RuO$_4$, this simple picture has to be necessarily modified. First of all most of these new superconductors are unconventional 
and nodal\cite{dsc1,revmod,annalen}. For more recent developments on this subject the reader may consult Ref. \cite{brazil}.

Parallel to this development, intense research has been done during the past few years in order to explore the 
properties of density wave with order parameter $\Delta(\bf k)$, which depends on the quasiparticle momentum along the Fermi surface. 
We call these states unconventional density waves (UDW) in parallel to unconventional superconductivity.

This kind of condensates was first speculated on by Halperin and Rice\cite{HR} as a possible ground state in the excitonic insulator. 
However 
unlike conventional density waves, there will be no x-ray signal or spin signal associated with UDW since the average of $\Delta(\bf 
k)$ over the Fermi surface usually vanishes (i.e. $\langle\Delta({\bf k})\rangle=0$).

Therefore one may think that UDW has a truly quantum mechanical order parameter somewhat similar to superconductors. 
UDW is not accompanied by the spatial variation of charge or spin. This intriguing property is known as hidden-order in recent 
literature\cite{nayak}. 
In order to make the hidden order visible, we need impurities for example\cite{roma}.
Also unlike conventional density waves the nodal excitations persist to $T=0$~K, giving rise to electronic specific heat $\sim T^2$, 
where $T$ is temperature. Indeed the thermodynamics is practically the same as the one for d-wave 
superconductors\cite{nagycikk,d-wave}. The recent surge in 
UDW is generated by the possibility that the pseudogap phase in high $T_c$ cuprates is d-wave density wave (d-DW)\cite{nayak,benfatto, 
app}. The angle resolved 
 photoemission spectra (ARPES) in the pseudogap phase indicate that the energy gap $\Delta(\bf k)$ is the same as in d-wave 
superconductors\cite{timusk}. Further the mysterious relation $\Delta(0)=2.14 T^*$ found in LSCO, YBCO and 
Bi2212\cite{oda,nakano,renner} can be readily interpreted in 
terms of d-DW. Here $\Delta(0)$ is the maximum value of the energy gap determined by STM and $T^*$ is the pseudogap temperature which 
is identified with the transition to  
d-DW. Actually $2.14$ is the weak-coupling  value for the d-wave superconductors.

The nature of the low-temperature phase (LTP) in quasi-two dimensional organic conductors 
$\alpha$-(BEDT-TTF)$_2$MHg(SCN)$_4$ with M=K, Rb and Tl has not been understood until recently\cite{singl}. Although the phase 
transition is 
clearly seen in magnetotransport measurements, neither charge, nor magnetic order has been established\cite{jetp,karts1}. Moreover the 
destruction of 
this LTP in an applied magnetic field suggests a kind of CDW rather than SDW.
On the other hand the temperature dependence of the threshold electric field associated with the sliding motion of DW\cite{ltp} is very 
different 
from the one in typical CDW but somewhat similar to the one in SDW\cite{tmtsf}. In fact, we have succeeded in describing the 
temperature dependence 
of the threshold electric field in terms of UCDW with imperfect nesting\cite{rapid,tesla}.

However, the LTP of $\alpha$-(BEDT-TTF)$_2$KHg(SCN)$_4$ is also well-known for its striking angular dependent magnetoresistance 
(ADMR)\cite{fermi,kovalev,caulfield2,hanasaki}. 
There have been many attempts to interpret this phenomenon in terms of the reconstructed Fermi surface. Rather we find that the Landau 
quantization of the quasiparticle orbit in UDW as described by Nersesyan et al.\cite{Ner1,Ner2} plays the crucial role 
here\cite{alfa,prl}. 
More recently we find that the same Landau quantization gives rise to large negative Nernst effect in UDW\cite{nernst}. Indeed we can 
describe the 
large Nernst effect observed in $\alpha$-(BEDT-TTF)$_2$KHg(SCN)$_4$\cite{choi} in terms of UCDW. Therefore we may conclude that the LTP 
in 
$\alpha$-(BEDT-TTF)$_2$KHg(SCN)$_4$ with M=K, Rb and Tl is UCDW. Also the LTP in $\alpha$-(BEDT-TTF)$_2$I$_3$ below $T_c=135$~K share 
many features common to UCDW\cite{dressel}. We shall discuss this briefly in Section 2.

The possibility of UCDW in 2H-NbSe$_2$ and USDW in the antiferromagnetic phase in URu$_2$Si$_2$ have also been 
suggested\cite{castroneto,IO}. We 
believe that the 
large negative Nernst effect observed in 2H-NbSe$_2$\cite{bel1} and the micromagnetism seen in URu$_2$Si$_2$\cite{roma,amitsuka} 
appeared to have confirmed UCDW in the former, 
USDW in the latter. As has already been mentioned, the pseudogap phase in high $T_c$ cuprates is most likely d-DW, though we prefer 
d-SDW to d-CDW\cite{app}.

In the following we shall first summarize the quasiparticle spectrum, the thermodynamics and other properties of UDW in Section 2. 
Then in Section 3.-5., we discuss  the Nersesyan effect for UDW in a magnetic field. This important work appears to be neglected by 
most people working on UDW. In particular the striking ADMR and the large negative Nernst signal immediately follow from the 
Nersesyan effect. Therefore in particular the giant Nernst effect is the hallmark of UDW.
We believe, that the large Nernst signal observed in underdoped LSCO, YBCO and Bi2212 indicates clearly that they are 
UDW\cite{capan,wang1,wang2,korea}.

In Section 6. we speculate the likely places where one can find UDW. Still very few UDW's have been identified. Therefore the field of 
UDW is still widely open and UDW will be found in unexpected places.

\section{BCS theory of unconventional density waves}

\begin{figure}[h!]
\psfrag{x}[t][b][1][0]{$E/\Delta$}
\psfrag{y}[b][t][1][0]{$N(E)/N_0$}
\centering\includegraphics[width=6cm,height=6cm]{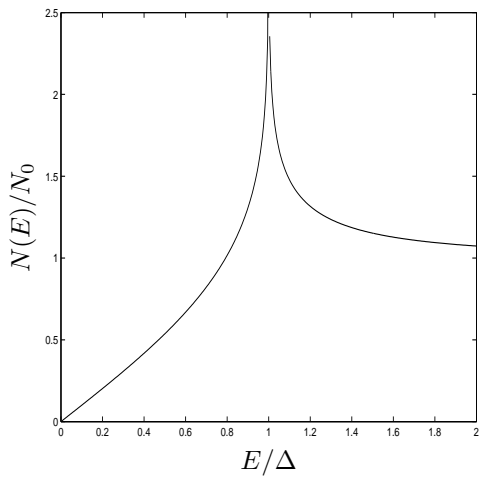}
\caption{The quasiparticle density of states of UDW is shown.}
\label{dos}
\end{figure}

In the following we shall consider quasi-one or quasi-two dimensional systems, with the Hamiltonian given by
\begin{equation}
H=\sum_{\bf k,\sigma}\xi({\bf k})a_{\bf k,\sigma}^{+}a_{\bf
 k,\sigma}+ \frac{1}{2} \sum_{\begin{array}{c}
          {\bf k,k^\prime,q} \\
          \sigma,\sigma^\prime
         \end{array}} {V}({\bf k,k^\prime,q})a_{\bf k+q,\sigma}^{+}
         a_{\bf k,\sigma}a_{\bf k^\prime-q,\sigma^\prime}^{+}a_{\bf
         k^\prime,\sigma^\prime},\label{ham}
\end{equation}
where $a_{\bf k,\sigma}^+$  and
 $a_{\bf k,\sigma}$ are the creation and annihilation operators of electrons with momentum $\bf k$ and spin $\sigma$, $\xi({\bf k})$ is 
the kinetic energy of electrons measured from the Fermi energy in the normal state and $V(\bf k,k^\prime,q)$ is the interaction between 
particles. In the following we shall approximate it as
\begin{equation}
V({\bf k,k^\prime,q})=2Vf({\bf k})f({\bf k^\prime})\delta(\bf q-Q),
\end{equation}
where $\bf Q$ is the nesting vector. Also for simplicity we limit ourselves to UCDW though a parallel treatment of USDW is possible.
Then within the mean field approximation Eq. (\ref{ham}) is recasted as 
\begin{equation}
H=\sum_{\bf k,\sigma}\left(\xi({\bf k})a_{\bf k,\sigma}^{+}a_{\bf k,\sigma}+\Delta({\bf k})a_{\bf k,\sigma}^{+}a_{\bf 
k+Q,\sigma}+\overline{\Delta}({\bf k})a_{\bf k+Q,\sigma}^{+}a_{\bf k,\sigma}\right)-\sum_{\bf k}\frac{|\Delta({\bf 
k})|^2}{V\langle|f({\bf k})|^2\rangle}
\end{equation}
and
\begin{equation}
\Delta({\bf k})=Vf({\bf k})\sum_{\bf k^\prime,\sigma}f({\bf k^\prime})\langle a_{\bf k^\prime-Q,\sigma}^{+}a_{\bf 
k^\prime,\sigma}\rangle.
\end{equation}
This is expressed in terms of Nambu's spinor\cite{nambu} as 
\begin{equation}
H=\sum_{\bf k,\sigma}\Psi^+_\sigma({\bf k})\left(\tilde\xi({\bf k})\rho_3+\eta({\bf k})+\Delta({\bf k})\rho_1\right)\Psi_\sigma(\bf 
k),
\end{equation}
where $\tilde\xi({\bf k})=(\xi({\bf k})-\xi({\bf k-Q}))/2$ and $\eta({\bf k})=(\xi({\bf k})+\xi({\bf k-Q}))/2$. In the following we 
shall take the tilde off from $\xi$. The Green's function is then given by
\begin{equation}
G^{-1}(\omega,{\bf k})=\omega-\xi({\bf k})\rho_3-\eta({\bf k})-\Delta({\bf k})\rho_1.
\label{green}
\end{equation}
The pole of $G(\omega,{\bf k})$ gives the quasiparticle spectrum as
\begin{equation}
\omega=\eta({\bf k})\pm\sqrt{\xi({\bf k})^2+\Delta({\bf k})^2}
\end{equation}
In most of quasi-one or quasi-two dimensional systems $\xi({\bf k})$ depends only on $\bf k$ perpendicular to the Fermi surface. So we 
can write $\xi({\bf k})=v(k_a-k_F)$, where $v$ is the Fermi velocity. Then in many cases we can take $\Delta({\bf k})=\Delta f({\bf 
k})$ with $f({\bf k})=\sin(bk_b)$ or $\cos(bk_b)$. Further if we can neglect $\eta({\bf k})$, the imperfect nesting term, the 
quasiparticle density of states is obtained as\cite{nagycikk}
\begin{eqnarray}
\frac{N(E)}{N_0}=\textmd{Re}|E|\left\langle\frac{1}{\sqrt{E^2-\Delta({\bf k})^2}}\right\rangle
=\left\{\begin{array}{l}
\frac 2\pi x K(x) \textmd{ for } x<1\\
\frac 2\pi  K(x^{-1}) \textmd{ for } x>1
\end{array}
\right.,
\end{eqnarray}
where $x=|E|/\Delta$ and $K(x)$ is the complete elliptic integral of the first kind. The quasiparticle density of states is shown in 
Fig. \ref{dos}. 

Also the thermodynamics is similarly obtained\cite{impurd-wave,hotta,universal}. For this purpose it is necessary to solve the gap 
equation
\begin{equation}
\lambda^{-1}=\langle f^2\rangle^{-1}\int_0^{vk_F}dE\textmd{Re}\langle\frac{f^2}{\sqrt{E^2-\Delta^2 f^2}}\rangle\tanh\frac{E}{2T}.
\end{equation}
Here we have neglected the imperfect nesting term for simplicity. The gap equation is the same as for d-wave superconductors. We obtain 
$\Delta(0)/T_c=2.14$ and
\begin{equation}
\frac{\Delta(T)}{\Delta(0)}\cong \sqrt{1-\left(\frac{T}{T_c}\right)^3}.
\label{approx}
\end{equation}

\begin{figure}[h!]
\psfrag{x}[t][b][1][0]{$T/T_c$}
\psfrag{y}[b][t][1][0]{$\Delta(T)/\Delta(0)$}
\centering\includegraphics[width=6cm,height=6cm]{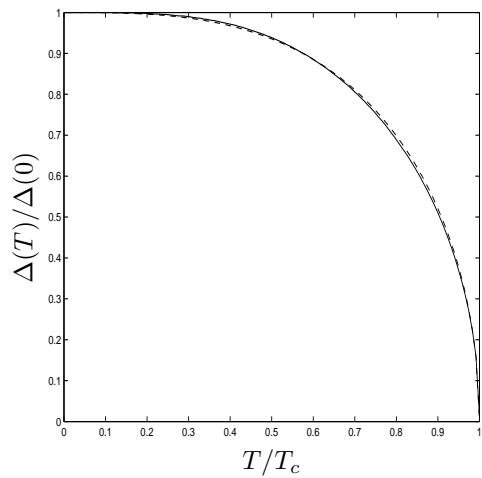}
\caption{The order parameter $\Delta(T)$ is shown (solid line) together with the approximate solution Eq. (\ref{approx}) 
(dashed line).}
\label{delta}
\end{figure}

In Fig. \ref{delta} the exact solution of $\Delta(T)$\cite{nagycikk} with the approximate one is shown. Therefore Eq. (\ref{approx}) is 
very useful 
for 
semiquantitative analysis. In very clean systems when the quasiparticle scattering is limited by impurity scattering, the electric 
conductivity is well approximated by
\begin{eqnarray}
\frac{\sigma(T)}{\sigma_n}=\frac 4\pi\int\limits_0^{\pi/2}d\phi \left(1+\exp{\beta\Delta \sin\phi}\right)^{-1}
\simeq\left\{\begin{array}{l}
1-\beta\Delta/\pi \textmd{ for } T\simeq T_c\\
\frac 4\pi\ln(2)(\beta\Delta)^{-1} \textmd{ for } T\ll T_c
\end{array}
\right..
\label{linear}
\end{eqnarray}
Unfortunately this $T$ linear behaviour for small temperatures cannot describe the $T^3$ dependence of the electric conductivity of 
$\alpha$-(BEDT-TTF)$_2$I$_3$\cite{dressel}. This suggests rather that the electric current $\bf J\parallel b$ is perpendicular to the 
nodal lines 
of UCDW. In other words this implies $\Delta({\bf k})=\Delta\sin(bk_b)$ in $\alpha$-(BEDT-TTF)$_2$I$_3$. This is somewhat surprising, 
since in $\alpha$-(BEDT-TTF)$_2$MHg(SCN)$_4$, what we are going to describe in some details, we found $\Delta({\bf 
k})=\Delta\sin(ck_c)$\cite{alfa,prl}. In the present configuration, we obtain
\begin{eqnarray}
\frac{\sigma_b(T)}{\sigma_{bn}}=\frac 8\pi\int\limits_0^{\pi/2}d\phi \sin(\phi)^2\left(1+\exp{\beta\Delta 
\sin\phi}\right)^{-1}\
\simeq\left\{\begin{array}{l}
1-\beta\Delta/3\pi \textmd{ for } T\simeq T_c\\
\frac {12}{\pi}\zeta(3)(\beta\Delta)^{-3} \textmd{ for } T\ll T_c
\end{array}
\right..
\end{eqnarray}
Especially for $T<T_c/2$, $\sigma_b(T)\sim T^3$, in accordance with the experimental data from $\alpha$-(BEDT-TTF)$_2$I$_3$.

\begin{figure}[h!]
\psfrag{x}[t][b][1][0]{$\omega$ [cm$^{-1}$]}
\psfrag{y}[b][t][1][0]{$\sigma(\omega)$ [($\Omega$cm)$^{-1}$]}
\centering\includegraphics[width=8cm,height=7cm]{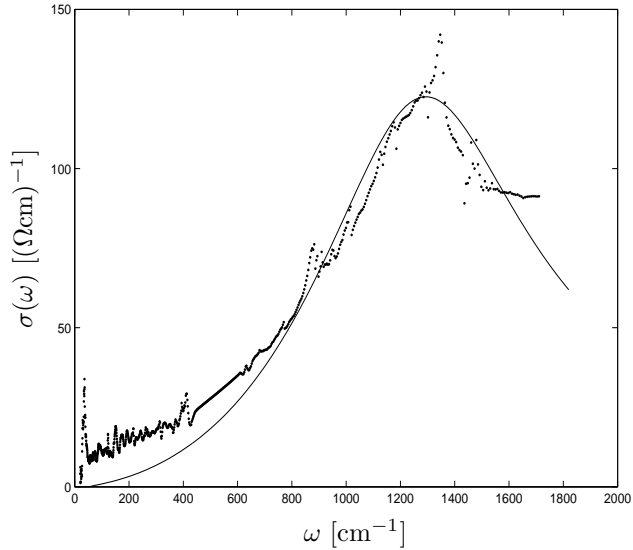}
\caption{The optical conductivity of $\alpha$-(BEDT-TTF)$_2$I$_3$ at $T=60$~K\cite{dressel} is plotted together with our theoretical 
prediction.}
\label{optcon}
\end{figure}

We show the optical conductivity with small impurity scattering rate in the Born limit\cite{scatter1,scatter2}, and the result is 
compared to the optical 
data taken at $T=60$~K in Fig. (\ref{optcon}). We think that the agreement is excellent. Also from this figure we can extract 
$\Delta\simeq 930$~K, which is 
about 3 times larger than the one expected from the weak-coupling theory $2.14\times 135\cong 290$~K. In conventional CDW and SDW, such 
a large deviation from the weak coupling theory result is mostly ascribed to the imperfect nesting term\cite{yamaji1,huang2}.

Therefore the transition in UDW is metal to metal. In general the conductivity in UDW is anisotropic reflecting the direction of $\bf 
J$ relative to the direction of the nodal lines. For example we believe that the $T$ linear resistivity in the pseudogap phase 
in high 
$T_c$ cuprates and heavy fermion systems are in part due to UDW. In these systems, the normal state resistance is proportional to $T^2$ 
as in the Landau Fermi liquid theory\cite{landau1,landau2,landau3}. In these systems, the quasiparticle lifetime is dominated by 
electron-electron scattering. Then 
in UDW this $T^2$ behaviour changes into $T$ linear resistance as is readily seen from Eq. (\ref{linear}). 
This behaviour is often called "non Fermi liquid". But we think that this word is very misleading and should only be used with care.
Actually the quasiparticle in UDW is bona fide Fermi particle. In the spirit of Landau, the Fermi liquid has to be defined as the 
Fermion, which has charge $\pm e$ and spin $1/2$ and is described by a pole of Green's function as given in Eq. (\ref{green}).
In this case, we can describe both the thermodynamics and transport properties of the system in terms of standard many body technique 
as in the book of Abrikosov, Gor'kov and Dzyaloshinskii\cite{klasszikus}.

\section{The Nersesyan effect}

This surprising effect of the magnetic field on the quasiparticle spectrum in UDW was first discussed in Ref. \cite{Ner1,Ner2}.
The quasiparticle spectrum in the presence of magnetic field is obtained from
\begin{equation}
\left(E-\xi({\bf k}+e{\bf A})\rho_3-\eta({\bf k}+e{\bf A})-\Delta({\bf k}+e{\bf A})\rho_1\right)\Psi({\bf r})=0,
\label{dirac}
\end{equation}
where we have introduced the magnetic field through the vector potential $\bf A$. It is readily recognized that Eq. (\ref{dirac}) has 
the same mathematical structure as the Dirac equation in a magnetic field studied in 1936\cite{heisenberg,weisskopf}.
For simplicity let us assume that the Fermi surface is parallel to the $a-c$ plane and the $b$ direction is perpendicular to the $a-c$ 
plane. Also $\xi(\bf k)$ depends only on $k_a$ while $\Delta(\bf k)$ only on $k_c$. Also for a while we neglect $\eta({\bf k})$ since 
in many cases $\eta({\bf k})\ll \max|\Delta({\bf k})|$.
Then the quasiparticle energy spectrum depends only on the magnetic field component parallel to $b$. It is more convenient to 
rewrite Eq. 
(\ref{dirac}) as\cite{nernst}
\begin{equation}
E\Psi=(-iv_a\partial_x\rho_3+\Delta ceB x \cos(\theta)\rho_1)\Psi,
\label{alap}
\end{equation}
$\theta$ is the angle the magnetic
field makes with the $b$ axis. 
We find
\begin{equation}
E^2=2nv_a\Delta c e |B\cos\theta|,
\end{equation}
where $n=0$, $1$, $2$\dots.
The Landau wavefunctions are given by
\begin{eqnarray}
\Psi_0=\left( \begin{array}{c}
i\\
1
\end{array}\right)\phi_0, \\
\Psi_{n\neq 0}=\frac{1}{\sqrt{2}}\left[\left(\begin{array}{c}
1\\
i
\end{array}\right)\phi_{n-1}\pm
\left(\begin{array}{c}
i\\
1
\end{array}\right)\phi_{n}\right],
\label{hullamfgv}
\end{eqnarray}
where $\phi_n$ is the $n$-th wavefunction of a linear harmonic oscillator with parameters
"mass" $m=1/2v_a^2$ and "frequency" $\omega=2 v_a \Delta ceB\cos(\theta)$. From Eq.
(\ref{hullamfgv}) it is obvious, that the $n\neq 0$ levels are
twofold degenerate, since $\Psi_{n\neq 0}$ is
composed of
the
$n-1$-th and $n$-th wavefunction of the harmonic oscillator.

So far we have neglected the imperfect nesting term. To be concrete, we assume that
\begin{equation}
\eta({\bf k})=\sum_{n=-\infty}^\infty \varepsilon_n\cos(2{\bf d}_n{\bf k}),
\end{equation}
where ${\bf d}_n$'s are selected lattice vectors. In the LTP of 
$\alpha$-(BEDT-TTF)$_2$KHg(SCN)$_4$, the multidip structure of the angular dependent magnetoresistance (ADMR) is accounted for by 
similar $\eta({\bf k})$\cite{prl}.
Then the imperfect nesting term removes the degeneracy of $E_{n\neq 0}$ and the Landau levels become
\begin{eqnarray}
E_{0,1}=-E_0^{(1)},\\
E_{1,1}=\pm E_1-E_1^{(1)},\\
E_{1,2}=\pm E_1-E_1^{(2)},
\end{eqnarray}
 and
\begin{eqnarray}
E_n=\sqrt{2nv_a\Delta ceB |\cos(\theta)|},\\
E_0^{(1)}=E_1^{(1)}=\sum_m\varepsilon_m\exp(-y_m),\\
E_1^{(2)}=\sum_m\varepsilon_m(1-2y_m)\exp(-y_m),
\end{eqnarray}
and $y_m=v_a {b}^2e |B\cos(\theta)|(\tan(\theta)\cos(\phi-\phi_o)-\tan(\theta_m))^2/\Delta c$, $\tan\theta_0\simeq 0.5$, 
$d_0\simeq 1.25$, $\phi_0\simeq 27^\circ$.
Here $\phi$ is the angle  the projected magnetic field on the $a$-$c$ plane makes
with the $c$-axis.

With the help of the quasiparticle spectrum, the thermodynamic properties are readily determined as done in Ref. \cite{Ner1,Ner2}.
In the following we shall consider the ADMR and the magnetothermopower.

\section{Angular dependent magnetoresistance (ADMR)}

\begin{figure}[h!]
\centering\psfrag{a}[t][b][1][0]{$a$}
\psfrag{b}[][][1][0]{$c$}
{\includegraphics[width=5cm,height=5cm]{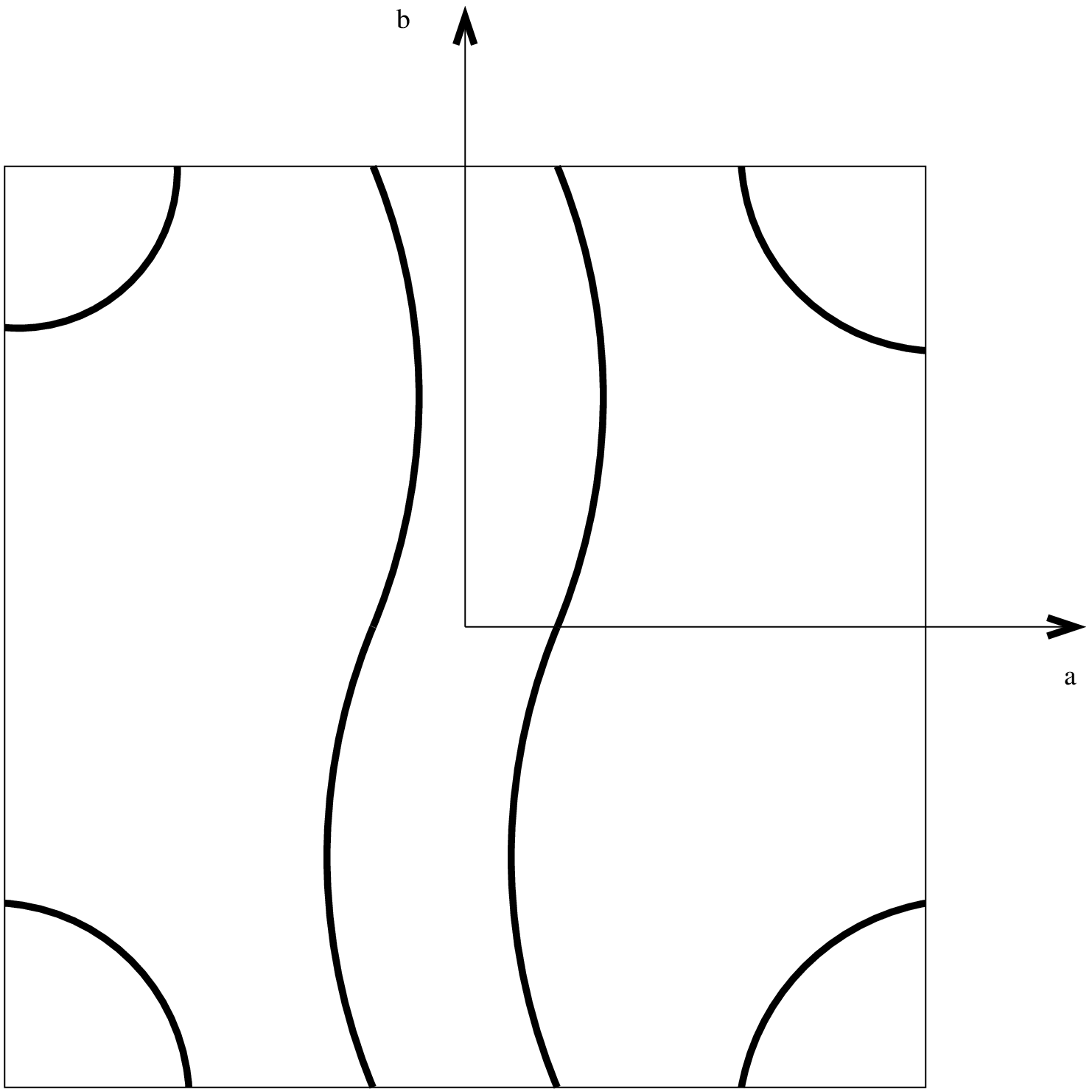}}
\label{fermisurf}
\hspace*{10mm}
\psfrag{B}[bl][tr][1][0]{$\bf B$}
\psfrag{c}[b][t][1][0]{$c$}
\psfrag{a}[b][t][1][0]{$a$}
\psfrag{b}[r][l][1][0]{$b$}
\psfrag{pp}[][][1][0]{$\phi$}
\psfrag{p}[][][1][0]{$\theta$}
\includegraphics[width=4cm,height=4cm]{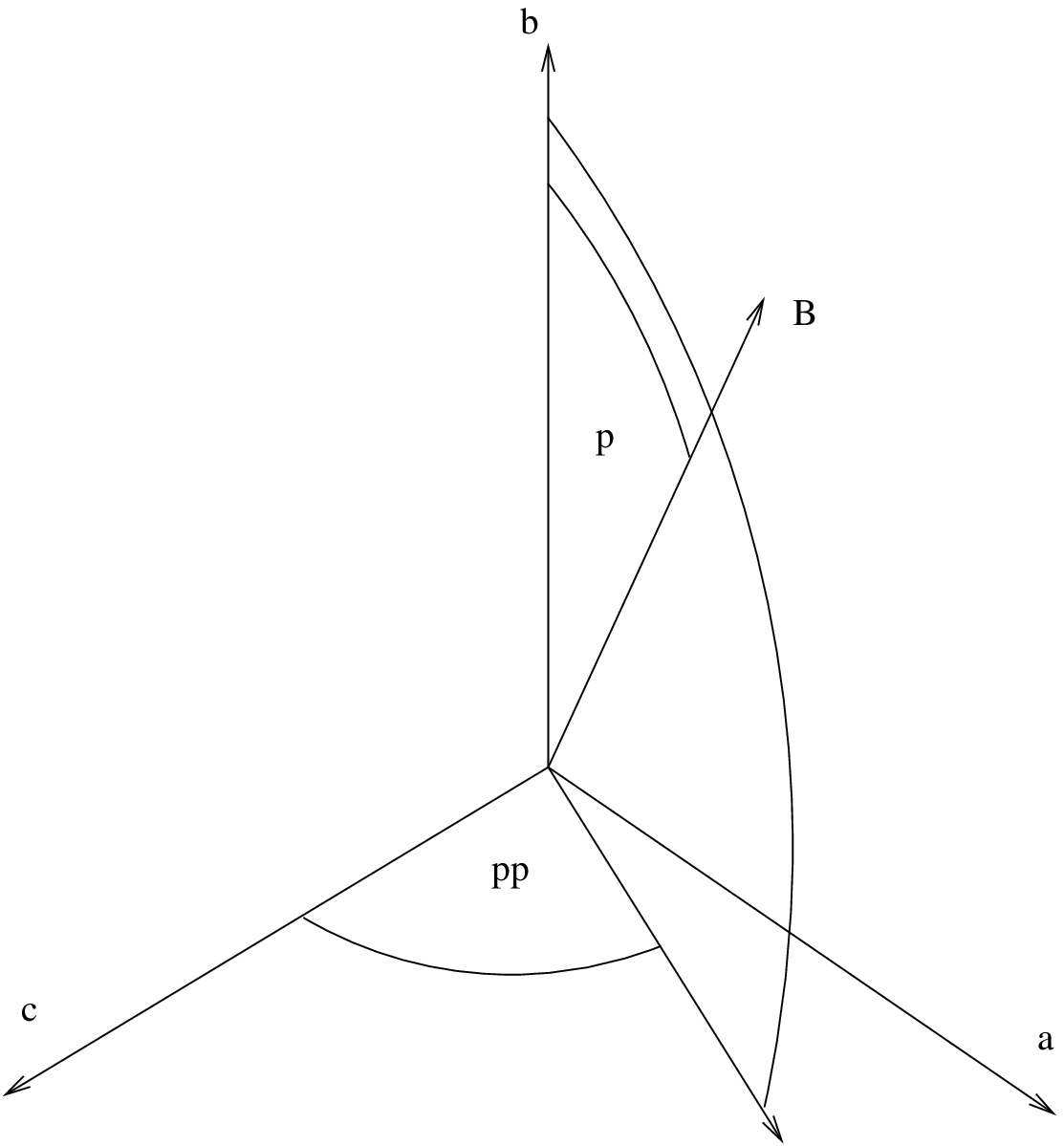}
\caption{The Fermi surface of $\alpha$-(BEDT-TTF)$_2$KHg(SCN)$_4$ is shown in the left panel. In the right one the
geometrical
configuration of the magnetic field with respect to
 the conducting plane is plotted.}
\end{figure}

In the low temperature and high field limit (i.e. $\beta E_1\gg 1$, where $\beta=1/T$), we assume
that the quasiparticle transport is dominated by the $n=0$ and $n=1$ Landau levels. Also for concreteness let us consider the 
magnetoresistance in the LTP of 
$\alpha$-(BEDT-TTF)$_2$MHg(SCN)$_4$ with M=K, Rb and Tl\cite{prl}.
The Fermi surface of this system is sketched in Fig. \ref{fermisurf} together with the related field configuration.

As is readily seen from Fig. \ref{fermisurf}, the Fermi surface in $\alpha$-(BEDT-TTF)$_2$MHg(SCN)$_4$ salts consists of quasi-one 
dimensional
sheets and quasi-two dimensional ellipses. Further we assume that UCDW appears only on the quasi-one dimensional sheets while the one
with the quasi-two dimensional Fermi surface remains in the normal state.
Then within the two level approximation, the magnetoresistance is given by
\begin{eqnarray}
R(B,\theta,\phi)^{-1}=2\sigma_1\left(\frac{\exp(-x_1)+\cosh(\zeta_0)}{\cosh(x_1)+\cosh(\zeta_0)}
+\frac{\exp(-x_1)+\cosh(\zeta_1)}{\cosh(x_1)+\cosh(\zeta_1)}\right)+\sigma_2,
\label{fit}
\end{eqnarray}
where $x_1=\beta E_1$, $\zeta_0=\beta E_1^{(1)}$, $\zeta_1=\beta E_1^{(2)}$. Here $\sigma_1$ and $\sigma_2$ are the conductivities
associated with the $n=1$ Landau level and the $n=0$ level plus the contribution from the
elliptical Fermi surface, respectively.
First let us consider the case when $B$ is normal to the conducting plane ($\theta=0$).
In the present configuration, the imperfect nesting plays no role and we can set $\zeta_0=\zeta_1=0$.

\begin{figure}[h]
\centering\psfrag{x}[t][b][1][0]{$B$(T)}
\psfrag{y}[b][t][1][0]{$R$(Ohm)}
\psfrag{m1}[t][b][1][0]{$T=1.4$K}
\psfrag{m2}[][][1][0]{$T=4.14$K}
\includegraphics[width=7cm,height=7cm]{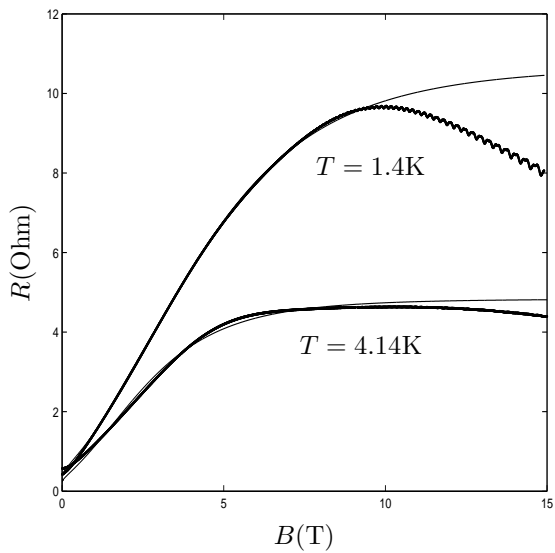}
\caption{The magnetoresistance is plotted for $T=1.4$K and $4.14$K as a function of magnetic field. The thick solid is the
experimental data, the thin one denotes our fit based on Eq. (\ref{fit}).}\label{koord}
\end{figure}

\begin{figure}[h]
\centering\psfrag{x}[t][b][1][0]{$T$(K)}
\psfrag{y}[b][t][1][0]{$R$(Ohm)}
\psfrag{m3}[l][r][1][0]{$B=15$T}
\includegraphics[width=7cm,height=7cm]{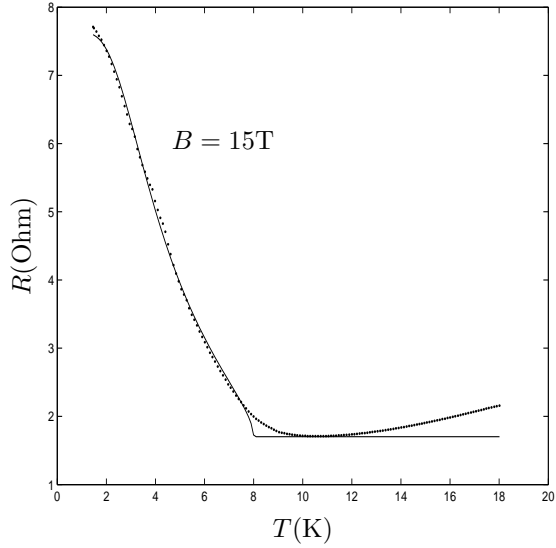}
\caption{The temperature dependent magnetoresistance is shown at $B=15$T. The dots are the experimental data, the
solid line is our fit.}\label{Rpara15T}
\end{figure}

In Fig. \ref{koord} we show the $B$ dependence of the magnetoresistance of single crystal of $\alpha$-(BEDT-TTF)$_2$KHg(SCN)$_4$ for 
$T=1.4$~K 
and $T=4.14$~K. As seen in Fig. \ref{koord}, the fitting improves further as the temperature decreases, though a clear deviation from 
the 
otherwise excellent fitting starts around $B=8$~T. This can come from the Landau quantization of the quasi-two dimensional Fermi 
surface, what we ignored so far. In Fig. \ref{Rpara15T},  we show the temperature dependence of the magnetoresistance for $B=15$~T. The 
fitting is 
almost perfect down to $T\simeq 2$~K. The deviation around $T=8$~K is clearly due to the fact that we have to include more Landau 
levels as $T$ approaches $T_c$. From these fittings we can extract $\sigma_2/\sigma_1\sim 0.1$ and $0.3$, $\Delta(0)=17$~K (the 
corresponding weak-coupling value), $v\sim 6\times 10^6$~cm/s. Also we have used Eq. (\ref{approx}) for $\Delta(T)$.

\begin{figure}[h]
\centering\psfrag{x}[t][b][1][0]{$\theta$ ($^\circ$)}
\psfrag{y}[b][t][1][0]{$R_\parallel(15T,\theta)$ (Ohm)}
\includegraphics[width=7cm,height=7cm]{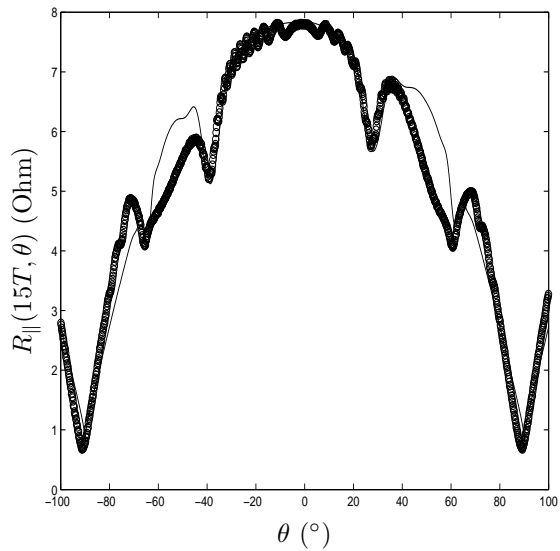}
\caption{The angular dependent magnetoresistance is shown for current parallel
to the a-c plane at $T=1.4$K,
$B=15$T. The open circles belong to the experimental data, the solid
line is our fit based on Eq. (\ref{fit}).}
\label{rpara}
\end{figure}

\begin{figure}[h]
\centering\psfrag{x}[t][b][1][0]{$\theta$ ($^\circ$)}
\psfrag{y}[b][t][1][0]{$R_\perp(15T,\theta)$ (Ohm)}
\includegraphics[width=7cm,height=7cm]{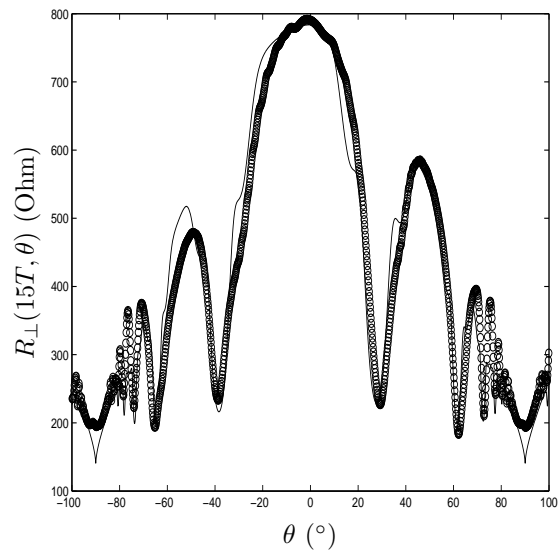}
\caption{The angular dependent magnetoresistance is shown for current perpendicular
to the a-c plane at $T=1.4$K,
$B=15$T. The open circles belong to the experimental data, the solid
line is our fit from Eq. (\ref{fit}).}
\label{rperp}
\end{figure}

\begin{figure}
\centering\psfrag{x}[t][b][1][0]{$\theta$ ($^\circ$)}
\psfrag{y}[b][t][1][0]{$R_\perp(15T,\theta)$ (Ohm)}
\includegraphics[width=6cm,height=6cm]{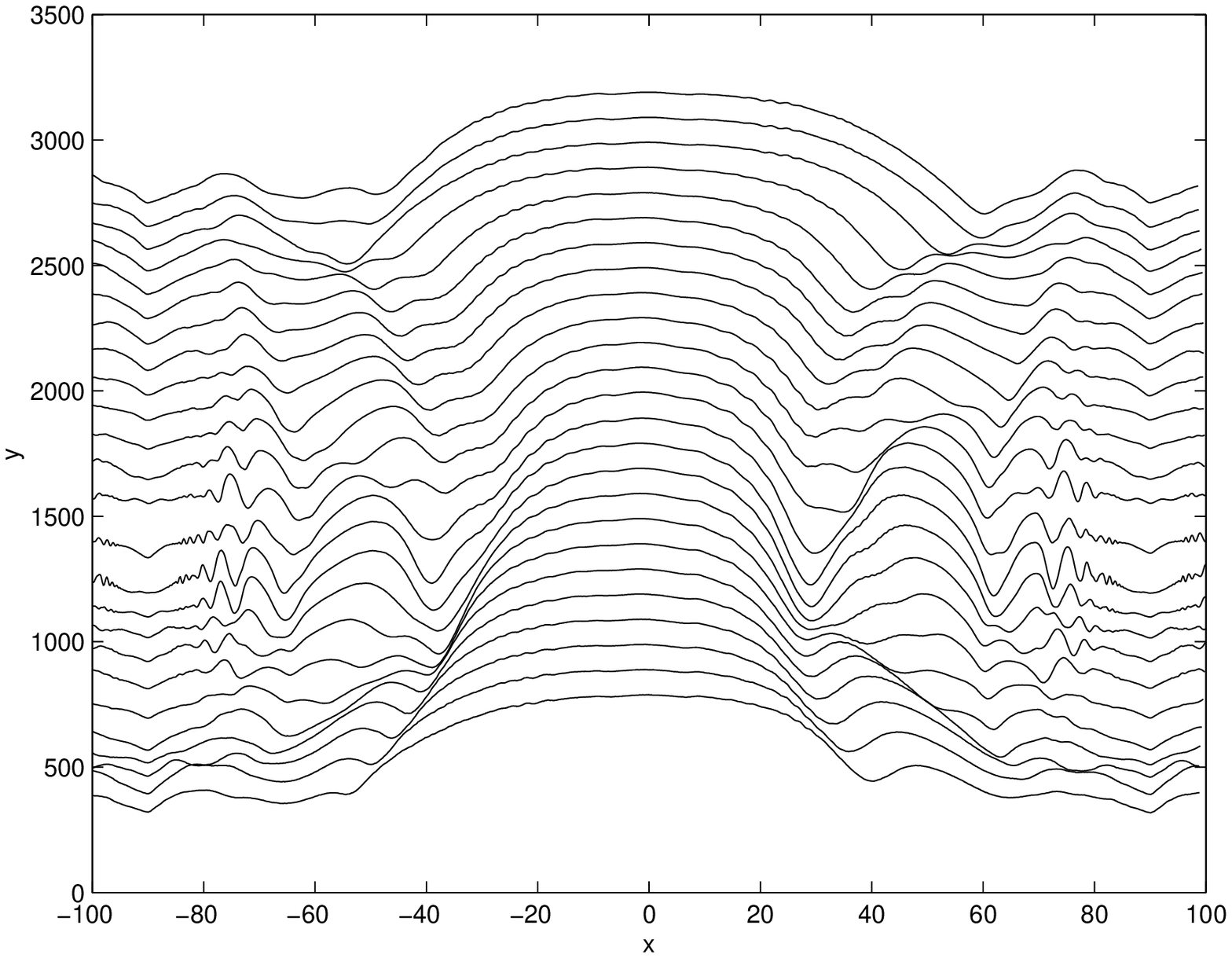}
\hspace*{2mm}
\psfrag{x}[t][b][1][0]{$\theta$ ($^\circ$)}
\psfrag{y}[b][t][1][0]{$R_\perp(15T,\theta)$ (Ohm)}
\includegraphics[width=6cm,height=6cm]{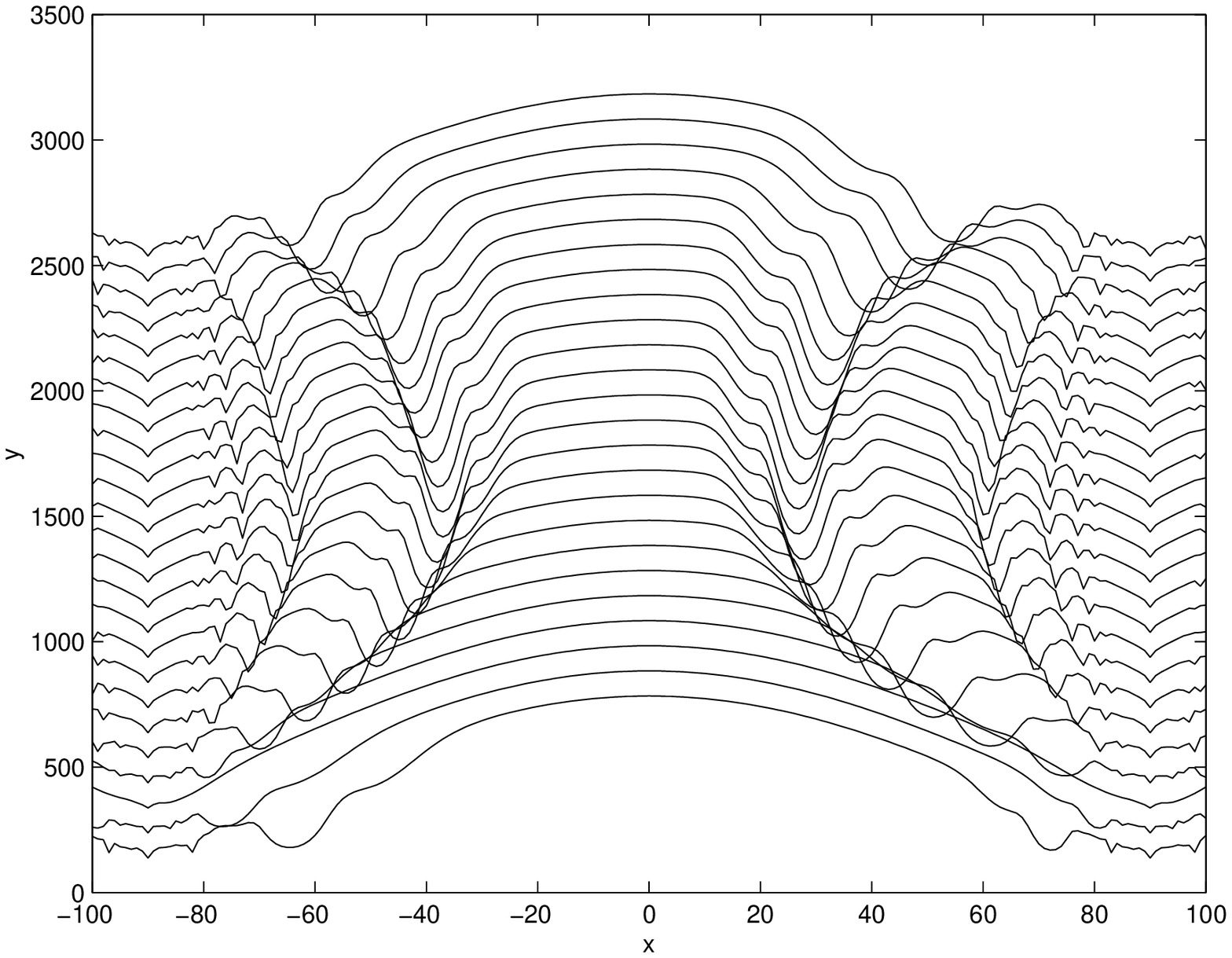}

\caption{ADMR is shown for current perpendicular to the a-c plane at $T=1.4$K and $B=15T$ for $\phi=-77^\circ$, $-70^\circ$,
$-62.5^\circ$, $-55^\circ$, $-47^\circ$, $-39^\circ$, $-30.5^\circ$, $-22^\circ$, $-14^\circ$, $-6^\circ$, $2^\circ$, $10^\circ$,
$23^\circ$, $33^\circ$, $41^\circ$, $48.5^\circ$, $56^\circ$, $61^\circ$, $64^\circ$,
$67^\circ$, $73^\circ$, $80^\circ$, $88.5^\circ$, $92^\circ$ and $96^\circ$ from bottom to top. The left (right) panel shows
experimental
(theoretical) curves, which are shifted from their original position along the vertical axis by $n\times100$Ohm, $n=0$ for
$\phi=-77^\circ$, $n=1$
for $\phi=-70^\circ$, \dots.}\label{sokphit}
\end{figure}

In Figs. \ref{rpara} and \ref{rperp}, we show the ADMR data taken at $T=1.4$~K, $B=15$~T and $\phi=45^\circ$.
As is readily seen, the fittings are excellent. From this we deduce  $\sigma_2/\sigma_1\sim 0.1$, $b\sim 30$~\AA, 
$\varepsilon_0\sim 3$~K
This $b$ is comparable to the lattice constant $b=20.26$~\AA. Finally in Fig. \ref{sokphit} we show $R$ versus $\theta$ for 
different $\phi$ 
and compare to the experimental data side by side. Perhaps there are still differences in some details, but the overall agreement is 
very striking. The present model can describe a similar figure found in Ref. \cite{hanasaki}  as well.
In summary, the Landau quantization of the quasiparticle spectrum in UDW as shown by Nersesyan et al.\cite{Ner1,Ner2} can account for 
the striking ADMR 
found in LTP of $\alpha$-(BEDT-TTF)$_2$KHg(SCN)$_4$. Very similar ADMR has been seen also in M=Rb and Tl compounds.
Therefore we conclude that LTP in $\alpha$-(BEDT-TTF)$_2$MHg(SCN)$_4$ should be UCDW. Also we believe that ADMR provides clear 
signature for the presence of UCDW or USDW.

Before closing this section, we note that very similar ADMR has been seen in Bechgaard salts (TMTSF)$_2$X with X=ClO$_4$, PF$_6$ and 
ReO$_4$, when the magnetic field is rotated within the $c^*-b$ plane\cite{osada,naughton,kang1,lee,kang2}. In particular, 
ADMR in PF$_6$ and ReO$_4$ compounds are very 
close to ADMR that we discussed so far. It is known that this striking angular dependence is seen only in the "normal state".
This means that conventional SDW and superconductivity has to be destroyed by pressure and magnetic field, respectively. Also
the magnetic field has to be less than the one which produces the field induced SDW\cite{ishiguro}.
Also the dips are called Lebed resonances\cite{lebed1,lebed2}, but the shape of ADMR has not been understood. Indeed, USDW in this 
$P-B$ 
phase diagram will 
describe this mysterious ADMR very consistently\cite{valami}.
We have suggested earlier, that USDW appears in (TMTSF)$_2$PF$_6$ in addition to conventional SDW for $T<T_c/3$, where $T_c$ is the 
transition temperature of SDW for $p<7$~kbar\cite{basleticprb}. Therefore the presence of USDW in a large area of the $P-B$ phase 
diagram may not be so 
surprising.

\section{Seebeck and Nernst effect}

The analysis in the preceding section is readily extended to the magnetothermopower tensor\cite{nernst}. First let us consider the 
diagonal 
magneto-thermoelectric power. This is given by
\begin{eqnarray}
S(B,\theta,\phi)=-\frac{R(B,\theta,\phi)k_B}{e}\left[\sigma_0 \zeta_0+\right.\nonumber\\
+\left.\sigma_1\left(\zeta_0
\frac{\exp(-x_1)+\cosh(\zeta_0)}{\cosh(x_1)+\cosh(\zeta_0)}+
\zeta_1 \frac{\exp(-x_1)+\cosh(\zeta_1)}{\cosh(x_1)+\cosh(\zeta_1)}+\right.\right.\nonumber \\
+\left.\left.x_1\left(
\frac{\sinh(\zeta_0)}{\cosh(x_1)+\cosh(\zeta_0)}+ \frac{\sinh(\zeta_1)}{\cosh(x_1)+\cosh(\zeta_1)}\right)\right)\right]
\label{tep}
\end{eqnarray}
where $x_1$, $\zeta_0$ and $\zeta_1$ have been defined after Eq. (\ref{fit}). In the thermoelectric 
power at low temperatures, the particle-hole symmetry breaking plays the crucial role. This means $S\sim \sigma_0$, $\sigma_1$.
In this treatment we have neglected the terms coming from $\partial\sigma/\partial\mu$, where $\mu$ is the chemical potential. For 
example in many heavy fermion systems, where the Kondo effect is apparent, the terms arising from $\partial\sigma/\partial\mu$ may be 
dominant. In this case we have
\begin{eqnarray}
S^K(B,\theta,\phi)=\frac{\pi^2k_B^2T}{3e} R(B,\theta,\phi)\times\nonumber \\
\times\left[2\frac{\partial\sigma_1}{\partial\mu}\left(
\frac{\exp(-x_1)+\cosh(\zeta_0)}{\cosh(x_1)+\cosh(\zeta_0)}+
 \frac{\exp(-x_1)+\cosh(\zeta_1)}{\cosh(x_1)+\cosh(\zeta_1)}\right)+
\frac{\partial\sigma_2}{\partial\mu}\right].
\end{eqnarray}
So in the most general case, these terms have to be added together.

\begin{figure}[h!]
\centering
\psfrag{x}[t][b][1.2][0]{$B$(T)}
\psfrag{y}[b][t][1.2][0]{$S$($\mu$V/K)}
\psfrag{x1}[l][r][1][0]{$T=1.4$~K}
\psfrag{x2}[r][l][1][0]{$T=4.8$~K}
\psfrag{x3}[l][][1][0]{$T=5.8$~K}
\psfrag{x4}[t][b][1][0]{$T=6.9$~K}
\includegraphics[width=7cm,height=7cm]{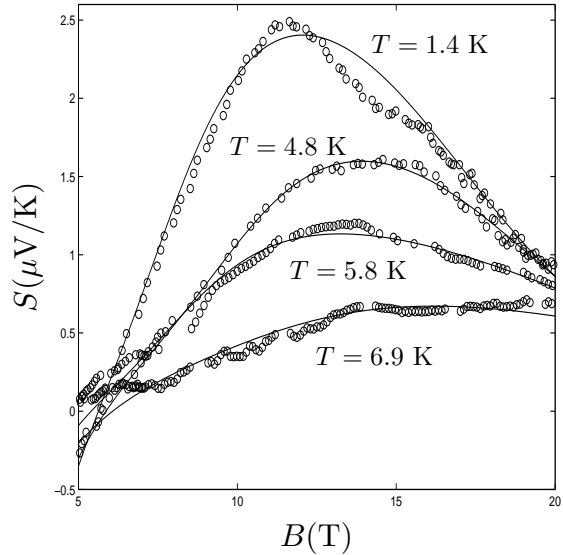}
\caption{The magnetothermopower for heat current along the $a$ direction is shown for $T=1.4$~K,
$T=4.8$~K ,$T=5.8$~K and $T=6.9$~K from top to bottom, the circles denote the experimental data from Ref.\cite{choi}, the solid line 
is our fit based on Eq. (\ref{tep})}.
\label{choi4a}
\end{figure}

\begin{figure}[h!]
\centering
\psfrag{x}[t][b][1.2][0]{$B$(T)}
\psfrag{y}[b][t][1.2][0]{$S_{xy}$($\mu$V/K)}
\psfrag{x6}[l][][1][0]{$T=1.4$~K}
\psfrag{x7}[l][][1][0]{$T=4.8$~K}
\includegraphics[width=7cm,height=7cm]{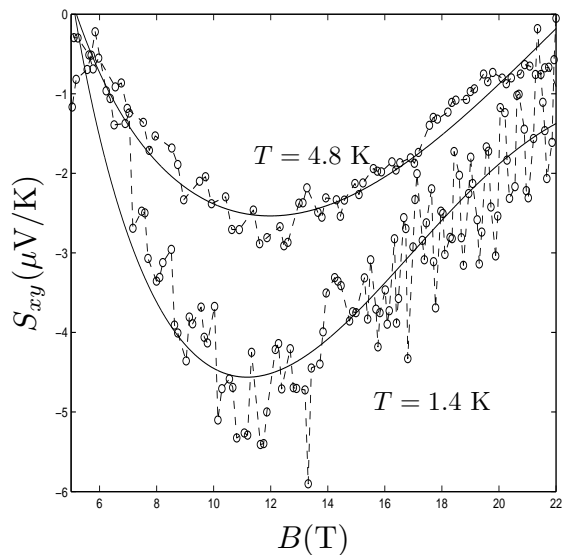}
\caption{The Nernst signal for heat current along the $a$ direction is shown
for $T=1.4$~K and  $T=4.8$~K (from bottom to top), the dashed lines with circles denote the
experimental data from Ref. \cite{choi}, the solid line is our fit based on Eq.
(\ref{nern}).}
\label{choi5a}
\end{figure}

\begin{figure}[h!]
\centering\psfrag{x}[t][b][1.2][0]{$T$(K)}
\psfrag{y}[b][t][1.2][0]{$S$($\mu$V/K)}
\psfrag{x8}[][][1][0]{$B=12$~T}
\includegraphics[width=7cm,height=7cm]{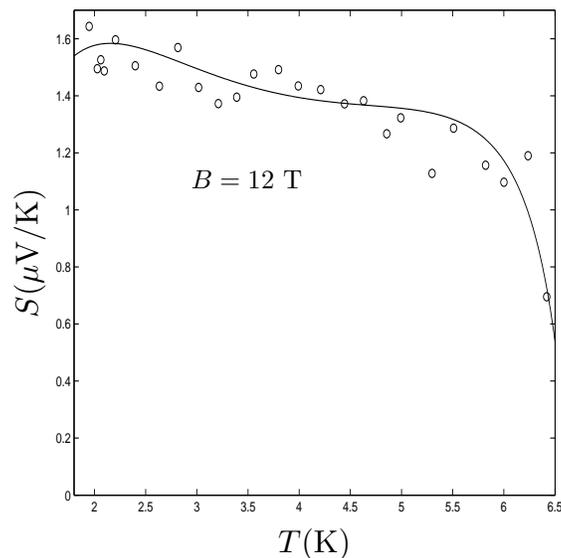}
\caption{The temperature dependence of the magnetothermopower for heat current along the $a$
direction is shown for $B=12$~T, the circles denote the experimental data from Ref. \cite{choi}, 
the solid line is our fit based on Eq. (\ref{tep}).}\label{choi6a}
\end{figure}

The Nernst effect is the off diagonal component of the thermoelectric power in the presence of
magnetic field. Also its formulation is different from above. We have seen already that
quasiparticle in UDW orbits around the magnetic field. Then when an electric field $\bf E$ is
applied within the conducting plane, the quasiparticle drifts with drift velocity ${\bf v}_D$ perpendicular to both 
$\bf B$ and $\bf E$ (${\bf v}_D=({\bf E\times B})/B^2$). Then the heat current parallel to ${\bf v}_D$ is
given by ${\bf J}_h=TS{\bf v}_D$, where $S$ is the entropy associated with the circling
quasiparticles
\begin{equation}
S=eB\sum_n\left[\ln(1+\exp(-\beta E_n))+\beta E_n(1+\exp(\beta
E_n))^{-1}\right],
\label{entr}
\end{equation}
the sum over $E_n$ has to be taken over all the Landau levels, and the magnetic field is assumed to be perpendicular to the $a-c$
plane ($\theta=0^\circ$). Also for simplicity we have neglected the imperfect nesting terms. Then for small $T$ and large $B$,
Eq. (\ref{entr}) is well approximated by taking the $n=0$ and $n=1$ Landau levels.
This gives
\begin{equation}
S=2eB\left[\ln(2)+2\ln\left(2\cosh\left(\frac{x_1}{2}\right)\right)
-x_1\tanh\left(\frac{x_1}{2}\right)\right].
\label{ner}
\end{equation}
So the Nernst coefficient in this configuration can be calculated, after considering the effect of the two dimensional parts
of the Fermi surface:
\begin{eqnarray}
S_{xy}=-\frac{S}{B\sigma}=\frac{1}{\sigma}\left[\frac{L_{\textmd{2D}}}{1+\gamma^2
B^2}-\right.\nonumber\\
\left.-2e\left(\ln(2)+2\ln\left(2\cosh\left(\frac{x_1}{2}\right)\right)
-x_1\tanh\left(\frac{x_1}{2}\right)\right)\right],
\label{nern}
\end{eqnarray}
where $\sigma=1/R=4\sigma_1/(\exp(x_1)+1)+\sigma_2$ from Eq. (\ref{fit}), $L_{\textmd{2D}}$ stems from the two dimensional cylinders
of
the Fermi
surface, $\gamma=e \tau/m$,
$\tau$ is the field-free relaxation time, $m$ is the effective mass of the electron.
Again $\sigma_1$ is related to the $n=1$ Landau level and $\sigma_2$ contains the contribution from the $n=0$ Landau levels as well
as from the elliptical Fermi surface.
Recently both the magneto-thermoelectric power and the Nernst effect of LTP in single crystal
$\alpha$-(BEDT-TTF)$_2$KHg(SCN)$_4$ has been reported\cite{choi}. We show in Figs. \ref{choi4a}, \ref{choi5a} and \ref{choi6a} the 
fitting of the 
experimental 
data with the theoretical 
expressions Eqs. (\ref{tep}) and (\ref{nern}). In these fittings again, we used
$\Delta\sim 17$~K, $v_a\sim 10^6$~cm/s and $\Delta(T)/\Delta(0)=\sqrt{1-(T/T_c)^3}$.
Both the Seebeck coefficient and the large negative Nernst effect are very consistently described in terms of UCDW\cite{nernst}.

As already mentioned, the large negative Nernst signal has been reported in the underdoped region of LSCO, YBCO and 
Bi2212\cite{capan,wang1,wang2}. A 
preliminary analysis indicated that the field dependence of this large Nernst effect can be described in terms of UDW\cite{korea}. Also 
it is well 
known that there are similarities between the 115 compounds CeCoIn$_5$ and high $T_c$ cuprate 
superconductors\cite{petrovic,izawa,shishida,sidorov}. These are quasi-two 
dimensionality, d$_{x^2-y^2}$-wave superconductivity and the proximity to antiferromagnetism.
More recently a large negative Nernst effect was observed above the superconducting transition temperature 
$T_c=2.3$~K\cite{bel2}. We 
believe that this indicates UDW above superconductivity in CeCoIn$_5$.

\section{Concluding remarks}

We have reviewed recent theoretical advances in understanding the physics of UDW (i.e. UCDW and USDW). This was in part stimulated
by the recent identification of the pseudogap phase of high $T_c$ cuprates as d-wave density wave (d-DW), and in part by the 
identification of LTP in $\alpha$-(BEDT-TTF)$_2$KHg(SCN)$_4$ as UCDW\cite{salerno}.

In particular in the latter case, the Landau quantization of the quasiparticle spectrum as discussed by Nersesyan et 
al.\cite{Ner1,Ner2} has played the 
crucial role. We propose here that both the striking angular dependent magnetoresistance (ADMR) and the large negative Nernst signal 
provide the hallmark of UDW.

For example, the peculiar ADMR seen in Bechgaard salts (TMTSF)$_2$PF$_6$ and (TMTSF)$_2$ReO$_4$ in the limited $P-B$ phase diagram 
suggests the presence of USDW in a wide region for $T<3$~K. Also UDW may inhabit many ground states in heavy fermion 
systems and organic conductors. In particular, the ground states in CeCoIn$_5$, URu$_2$Si$_2$, CeCu$_2$Si$_2$, 
$\kappa$-(BEDT-TTF)$_2$X with X=Cu(NCS)$_2$, Cu[N(CN)$_2$]Br, Cu[N(CN)$_2$]Cl should be explored 
further\cite{kanoda,izawa2,muller,pinteric}. Since the beginning of the 
21st century, the gap symmetry of unconventional superconductors Sr$_2$RuO$_4$, CeCoIn$_5$, $\kappa$-(BEDT-TTF)$_2$Cu(NCS)$_2$, 
YNi$_2$B$_2$C and PrOs$_4$Sb have been determined. Likewise UDW becomes the density wave of the 21st century. In spite of very limited 
findings on this newly developing subject, we are confident that new discovery and new understanding of truly quantum condensates will 
modify and enrich our perspective on condensed matter physics in general. The vast forest with exotic birds and flowers are waiting for 
our exploration.

\section*{Acknowledgments}

We thank Mario Basleti\'c, Stephan Haas, Koichi Izawa, Hae-Young Kee,  Bojana Korin-Hamzi\'c, Mark Kartsovnik, Yuji Matsuda, Marko 
Pinteri\'c, Silvia Tomi\'c, Peter Thalmeier and Hyekyung Won for collaborations and discussions on related subjects. Also one of us (K. 
M.) acknowledges gratefully the hospitality of Max-Planck Institute for the Physics of Complex Systems for every summer since 1995.
This work was supported by the Hungarian Scientific Research Fund under grant numbers OTKA T046269, NDF45172 and T037451.

\bibliographystyle{apsrev}
\bibliography{mplbcite}

\begin{thebibliography}{10}
\expandafter\ifx\csname bibnamefont\endcsname\relax
  \def\bibnamefont#1{#1}\fi
\expandafter\ifx\csname bibfnamefont\endcsname\relax
  \def\bibfnamefont#1{#1}\fi
\expandafter\ifx\csname url\endcsname\relax
  \def\url#1{\texttt{#1}}\fi
\expandafter\ifx\csname urlprefix\endcsname\relax\def\urlprefix{URL }\fi
\providecommand{\bibinfo}[2]{#2}
\providecommand{\eprint}[2][]{\url{#2}}

\bibitem{solyom}
\bibinfo{author}{\bibfnamefont{J.}~\bibnamefont{S{\'o}lyom}},
  \bibinfo{journal}{Adv. Phys.} \textbf{\bibinfo{volume}{28}},
  \bibinfo{pages}{201} (\bibinfo{year}{1979}).

\bibitem{jerome}
\bibinfo{author}{\bibfnamefont{D.}~\bibnamefont{Jerome}} \bibnamefont{and}
  \bibinfo{author}{\bibfnamefont{H.}~\bibnamefont{Schulz}},
  \bibinfo{journal}{Adv. Phys.} \textbf{\bibinfo{volume}{31}},
  \bibinfo{pages}{299} (\bibinfo{year}{1982}).

\bibitem{gruner}
\bibinfo{author}{\bibfnamefont{G.}~\bibnamefont{Gr\"uner}},
  \emph{\bibinfo{title}{Density waves in solids}}
  (\bibinfo{publisher}{Addison-Wesley}, \bibinfo{address}{Reading},
  \bibinfo{year}{1994}).

\bibitem{ishiguro}
\bibinfo{author}{\bibfnamefont{T.}~\bibnamefont{Ishiguro}},
  \bibinfo{author}{\bibfnamefont{K.}~\bibnamefont{Yamaji}}, \bibnamefont{and}
  \bibinfo{author}{\bibfnamefont{G.}~\bibnamefont{Saito}}
  (\bibinfo{publisher}{Springer}, \bibinfo{address}{Berlin},
  \bibinfo{year}{1999}).

\bibitem{BCS}
\bibinfo{author}{\bibfnamefont{J.}~\bibnamefont{Bardeen}},
  \bibinfo{author}{\bibfnamefont{L.~N.} \bibnamefont{Cooper}},
  \bibnamefont{and} \bibinfo{author}{\bibfnamefont{J.~R.}
  \bibnamefont{Schrieffer}}, \bibinfo{journal}{Phys. Rev.}
  \textbf{\bibinfo{volume}{108}}, \bibinfo{pages}{1175} (\bibinfo{year}{1957}).

\bibitem{dsc1}
\bibinfo{author}{\bibfnamefont{D.~J.~V.} \bibnamefont{Harlingen}},
  \bibinfo{journal}{Rev. Mod. Phys.} \textbf{\bibinfo{volume}{67}},
  \bibinfo{pages}{515} (\bibinfo{year}{1995}).

\bibitem{revmod}
\bibinfo{author}{\bibfnamefont{C.~C.} \bibnamefont{Tsuei}} \bibnamefont{and}
  \bibinfo{author}{\bibfnamefont{J.~R.} \bibnamefont{Kirtley}},
  \bibinfo{journal}{Rev. Mod.Phys.} \textbf{\bibinfo{volume}{72}},
  \bibinfo{pages}{969} (\bibinfo{year}{2000}).

\bibitem{annalen}
\bibinfo{author}{\bibfnamefont{K.}~\bibnamefont{Maki}} \bibnamefont{and}
  \bibinfo{author}{\bibfnamefont{H.}~\bibnamefont{Won}}, \bibinfo{journal}{Ann.
  Phys. (Leipzig)} \textbf{\bibinfo{volume}{5}}, \bibinfo{pages}{320}
  (\bibinfo{year}{1996}).

\bibitem{brazil}
\bibinfo{author}{\bibfnamefont{H.}~\bibnamefont{Won}},
  \bibinfo{author}{\bibfnamefont{Q.}~\bibnamefont{Yuan}},
  \bibinfo{author}{\bibfnamefont{P.}~\bibnamefont{Thalmeier}},
  \bibnamefont{and} \bibinfo{author}{\bibfnamefont{K.}~\bibnamefont{Maki}},
  \bibinfo{journal}{Braz. J. Phys.} \textbf{\bibinfo{volume}{33}},
  \bibinfo{pages}{675} (\bibinfo{year}{2003}).

\bibitem{HR}
\bibinfo{author}{\bibfnamefont{B.~I.} \bibnamefont{Halperin}} \bibnamefont{and}
  \bibinfo{author}{\bibfnamefont{T.~M.} \bibnamefont{Rice}}, in
  \emph{\bibinfo{booktitle}{Solid State Physics}}, edited by
  \bibinfo{editor}{\bibfnamefont{F.}~\bibnamefont{Seitz}},
  \bibinfo{editor}{\bibfnamefont{D.}~\bibnamefont{Turnbull}}, \bibnamefont{and}
  \bibinfo{editor}{\bibfnamefont{H.}~\bibnamefont{Ehrenreich}}
  (\bibinfo{publisher}{Academic Press}, \bibinfo{address}{New York},
  \bibinfo{year}{1968}), vol.~\bibinfo{volume}{21}, p. \bibinfo{pages}{115}.

\bibitem{nayak}
\bibinfo{author}{\bibfnamefont{S.}~\bibnamefont{Chakravarty}},
  \bibinfo{author}{\bibfnamefont{R.~B.} \bibnamefont{Laughlin}},
  \bibinfo{author}{\bibfnamefont{D.~K.} \bibnamefont{Morr}}, \bibnamefont{and}
  \bibinfo{author}{\bibfnamefont{C.}~\bibnamefont{Nayak}},
  \bibinfo{journal}{Phys. Rev. B} \textbf{\bibinfo{volume}{63}},
  \bibinfo{pages}{094503} (\bibinfo{year}{2001}).

\bibitem{roma}
\bibinfo{author}{\bibfnamefont{A.}~\bibnamefont{Virosztek}},
  \bibinfo{author}{\bibfnamefont{K.}~\bibnamefont{Maki}}, \bibnamefont{and}
  \bibinfo{author}{\bibfnamefont{B.}~\bibnamefont{D{\'o}ra}},
  \bibinfo{journal}{Int. J. Mod. Phys. B} \textbf{\bibinfo{volume}{16}},
  \bibinfo{pages}{1667} (\bibinfo{year}{2002}).

\bibitem{nagycikk}
\bibinfo{author}{\bibfnamefont{B.}~\bibnamefont{D{\'o}ra}} \bibnamefont{and}
  \bibinfo{author}{\bibfnamefont{A.}~\bibnamefont{Virosztek}},
  \bibinfo{journal}{Eur. Phys. J. B} \textbf{\bibinfo{volume}{22}},
  \bibinfo{pages}{167} (\bibinfo{year}{2001}).

\bibitem{d-wave}
\bibinfo{author}{\bibfnamefont{H.}~\bibnamefont{Won}} \bibnamefont{and}
  \bibinfo{author}{\bibfnamefont{K.}~\bibnamefont{Maki}},
  \bibinfo{journal}{Phys. Rev. B} \textbf{\bibinfo{volume}{49}},
  \bibinfo{pages}{1397} (\bibinfo{year}{1994}).

\bibitem{benfatto}
\bibinfo{author}{\bibfnamefont{L.}~\bibnamefont{Benfatto}},
  \bibinfo{author}{\bibfnamefont{S.}~\bibnamefont{Caprara}}, \bibnamefont{and}
  \bibinfo{author}{\bibfnamefont{C.}~\bibnamefont{{Di Castro}}},
  \bibinfo{journal}{Eur. Phys. J. B} \textbf{\bibinfo{volume}{17}},
  \bibinfo{pages}{95} (\bibinfo{year}{2000}).

\bibitem{app}
\bibinfo{author}{\bibfnamefont{B.}~\bibnamefont{D\'ora}},
  \bibinfo{author}{\bibfnamefont{A.}~\bibnamefont{Virosztek}},
  \bibnamefont{and} \bibinfo{author}{\bibfnamefont{K.}~\bibnamefont{Maki}},
  \bibinfo{journal}{Acta Physica Polonica B} \textbf{\bibinfo{volume}{34}},
  \bibinfo{pages}{571} (\bibinfo{year}{2003}).

\bibitem{timusk}
\bibinfo{author}{\bibfnamefont{T.}~\bibnamefont{Timusk}} \bibnamefont{and}
  \bibinfo{author}{\bibfnamefont{B.}~\bibnamefont{Statt}},
  \bibinfo{journal}{Rep. Prog. Phys.} \textbf{\bibinfo{volume}{62}},
  \bibinfo{pages}{61} (\bibinfo{year}{1999}).

\bibitem{oda}
\bibinfo{author}{\bibfnamefont{M.}~\bibnamefont{Oda}},
  \bibinfo{author}{\bibfnamefont{R.}~\bibnamefont{Kubota}},
  \bibinfo{author}{\bibfnamefont{K.}~\bibnamefont{Hoya}},
  \bibinfo{author}{\bibfnamefont{C.}~\bibnamefont{Manabe}},
  \bibinfo{author}{\bibfnamefont{N.}~\bibnamefont{Momono}},
  \bibinfo{author}{\bibfnamefont{T.}~\bibnamefont{Nakano}}, \bibnamefont{and}
  \bibinfo{author}{\bibfnamefont{M.}~\bibnamefont{Ido}},
  \bibinfo{journal}{Physica C} \textbf{\bibinfo{volume}{281}},
  \bibinfo{pages}{135} (\bibinfo{year}{1997}).

\bibitem{nakano}
\bibinfo{author}{\bibfnamefont{T.}~\bibnamefont{Nakano}},
  \bibinfo{author}{\bibfnamefont{N.}~\bibnamefont{Momono}},
  \bibinfo{author}{\bibfnamefont{M.}~\bibnamefont{Oda}}, \bibnamefont{and}
  \bibinfo{author}{\bibfnamefont{M.}~\bibnamefont{Ido}}, \bibinfo{journal}{J.
  Phys. Soc. Jpn} \textbf{\bibinfo{volume}{67}}, \bibinfo{pages}{2622}
  (\bibinfo{year}{1998}).

\bibitem{renner}
\bibinfo{author}{\bibfnamefont{M.}~\bibnamefont{Kugler}},
  \bibinfo{author}{\bibfnamefont{O.}~\bibnamefont{Fischer}},
  \bibinfo{author}{\bibnamefont{{Ch. Renner}}},
  \bibinfo{author}{\bibfnamefont{S.}~\bibnamefont{Ono}}, \bibnamefont{and}
  \bibinfo{author}{\bibfnamefont{Y.}~\bibnamefont{Ando}},
  \bibinfo{journal}{Phys. Rev. Lett.} \textbf{\bibinfo{volume}{86}},
  \bibinfo{pages}{4911} (\bibinfo{year}{2001}).

\bibitem{singl}
\bibinfo{author}{\bibfnamefont{J.}~\bibnamefont{Singleton}},
  \bibinfo{journal}{Rep. Prog. Phys.} \textbf{\bibinfo{volume}{63}},
  \bibinfo{pages}{1161} (\bibinfo{year}{2000}).

\bibitem{jetp}
\bibinfo{author}{\bibfnamefont{P.}~\bibnamefont{Christ}},
  \bibinfo{author}{\bibfnamefont{W.}~\bibnamefont{Biberacher}},
  \bibinfo{author}{\bibfnamefont{M.~V.} \bibnamefont{Kartsovnik}},
  \bibinfo{author}{\bibfnamefont{E.}~\bibnamefont{Steep}},
  \bibinfo{author}{\bibfnamefont{E.}~\bibnamefont{Balthes}},
  \bibinfo{author}{\bibfnamefont{H.}~\bibnamefont{Weiss}}, \bibnamefont{and}
  \bibinfo{author}{\bibfnamefont{H.}~\bibnamefont{M{\"u}ller}},
  \bibinfo{journal}{JETP Lett.} \textbf{\bibinfo{volume}{71}},
  \bibinfo{pages}{303} (\bibinfo{year}{2000}).

\bibitem{karts1}
\bibinfo{author}{\bibfnamefont{D.}~\bibnamefont{Andres}},
  \bibinfo{author}{\bibfnamefont{M.~V.} \bibnamefont{Kartsovnik}},
  \bibinfo{author}{\bibfnamefont{W.}~\bibnamefont{Biberacher}},
  \bibinfo{author}{\bibfnamefont{H.}~\bibnamefont{Weiss}},
  \bibinfo{author}{\bibfnamefont{E.}~\bibnamefont{Balthes}},
  \bibinfo{author}{\bibfnamefont{H.}~\bibnamefont{M{\"u}ller}},
  \bibnamefont{and} \bibinfo{author}{\bibfnamefont{N.}~\bibnamefont{Kushch}},
  \bibinfo{journal}{Phys. Rev. B} \textbf{\bibinfo{volume}{64}},
  \bibinfo{pages}{161104(R)} (\bibinfo{year}{2001}).

\bibitem{ltp}
\bibinfo{author}{\bibfnamefont{M.}~\bibnamefont{Basleti{\'c}}},
  \bibinfo{author}{\bibfnamefont{B.}~\bibnamefont{Korin-Hamzi{\'c}}},
  \bibinfo{author}{\bibfnamefont{M.~V.} \bibnamefont{Kartsovnik}},
  \bibnamefont{and}
  \bibinfo{author}{\bibfnamefont{H.}~\bibnamefont{M{\"u}ller}},
  \bibinfo{journal}{Synth. Met.} \textbf{\bibinfo{volume}{120}},
  \bibinfo{pages}{1021} (\bibinfo{year}{2001}).

\bibitem{tmtsf}
\bibinfo{author}{\bibnamefont{see for~example S.~Tomi{\'c}}},
  \bibinfo{author}{\bibfnamefont{J.~R.} \bibnamefont{Cooper}},
  \bibinfo{author}{\bibfnamefont{W.}~\bibnamefont{Kang}},
  \bibinfo{author}{\bibfnamefont{D.}~\bibnamefont{Jerome}}, \bibnamefont{and}
  \bibinfo{author}{\bibfnamefont{K.}~\bibnamefont{Maki}}, \bibinfo{journal}{J.
  Phys. I (Paris)} \textbf{\bibinfo{volume}{1}}, \bibinfo{pages}{1603}
  (\bibinfo{year}{1991}).

\bibitem{rapid}
\bibinfo{author}{\bibfnamefont{B.}~\bibnamefont{D\'ora}},
  \bibinfo{author}{\bibfnamefont{A.}~\bibnamefont{Virosztek}},
  \bibnamefont{and} \bibinfo{author}{\bibfnamefont{K.}~\bibnamefont{Maki}},
  \bibinfo{journal}{Phys. Rev. B} \textbf{\bibinfo{volume}{64}},
  \bibinfo{pages}{041101(R)} (\bibinfo{year}{2001}).

\bibitem{tesla}
\bibinfo{author}{\bibfnamefont{B.}~\bibnamefont{D\'ora}},
  \bibinfo{author}{\bibfnamefont{A.}~\bibnamefont{Virosztek}},
  \bibnamefont{and} \bibinfo{author}{\bibfnamefont{K.}~\bibnamefont{Maki}},
  \bibinfo{journal}{Phys. Rev. B} \textbf{\bibinfo{volume}{65}},
  \bibinfo{pages}{155119} (\bibinfo{year}{2002}).

\bibitem{fermi}
\bibinfo{author}{\bibfnamefont{T.}~\bibnamefont{Sasaki}} \bibnamefont{and}
  \bibinfo{author}{\bibfnamefont{N.}~\bibnamefont{Toyota}},
  \bibinfo{journal}{Phys. Rev. B} \textbf{\bibinfo{volume}{49}},
  \bibinfo{pages}{10120} (\bibinfo{year}{1994}).

\bibitem{kovalev}
\bibinfo{author}{\bibfnamefont{A.~E.} \bibnamefont{Kovalev}},
  \bibinfo{author}{\bibfnamefont{M.~V.} \bibnamefont{Kartsovnik}},
  \bibinfo{author}{\bibfnamefont{R.~P.} \bibnamefont{Shibaeva}},
  \bibinfo{author}{\bibfnamefont{L.~P.} \bibnamefont{Rozenberg}},
  \bibinfo{author}{\bibfnamefont{I.~F.} \bibnamefont{Schegolev}},
  \bibnamefont{and} \bibinfo{author}{\bibfnamefont{N.~D.}
  \bibnamefont{Kushch}}, \bibinfo{journal}{Solid State Commun.}
  \textbf{\bibinfo{volume}{89}}, \bibinfo{pages}{575} (\bibinfo{year}{1994}).

\bibitem{caulfield2}
\bibinfo{author}{\bibfnamefont{J.}~\bibnamefont{Caulfield}},
  \bibinfo{author}{\bibfnamefont{J.}~\bibnamefont{Singleton}},
  \bibinfo{author}{\bibfnamefont{P.~T.~J.} \bibnamefont{Hendriks}},
  \bibinfo{author}{\bibfnamefont{J.~A. A.~J.} \bibnamefont{Perenboom}},
  \bibinfo{author}{\bibfnamefont{F.~L.} \bibnamefont{Pratt}},
  \bibinfo{author}{\bibfnamefont{M.}~\bibnamefont{Doporto}},
  \bibinfo{author}{\bibfnamefont{W.}~\bibnamefont{Hayes}},
  \bibinfo{author}{\bibfnamefont{M.}~\bibnamefont{Kurmoo}}, \bibnamefont{and}
  \bibinfo{author}{\bibfnamefont{P.}~\bibnamefont{Day}}, \bibinfo{journal}{J.
  Phys. Cond. Mat.} \textbf{\bibinfo{volume}{6}}, \bibinfo{pages}{L155}
  (\bibinfo{year}{1994}).

\bibitem{hanasaki}
\bibinfo{author}{\bibfnamefont{N.}~\bibnamefont{Hanasaki}},
  \bibinfo{author}{\bibfnamefont{S.}~\bibnamefont{Kagoshima}},
  \bibinfo{author}{\bibfnamefont{N.}~\bibnamefont{Miura}}, \bibnamefont{and}
  \bibinfo{author}{\bibfnamefont{G.}~\bibnamefont{Saito}}, \bibinfo{journal}{J.
  Phys. Soc. Japan} \textbf{\bibinfo{volume}{65}}, \bibinfo{pages}{1010}
  (\bibinfo{year}{1996}).

\bibitem{Ner1}
\bibinfo{author}{\bibfnamefont{A.~A.} \bibnamefont{Nersesyan}}
  \bibnamefont{and} \bibinfo{author}{\bibfnamefont{G.~E.}
  \bibnamefont{Vachnadze}}, \bibinfo{journal}{J. Low T. Phys.}
  \textbf{\bibinfo{volume}{77}}, \bibinfo{pages}{293} (\bibinfo{year}{1989}).

\bibitem{Ner2}
\bibinfo{author}{\bibfnamefont{A.~A.} \bibnamefont{Nersesyan}},
  \bibinfo{author}{\bibfnamefont{G.~I.} \bibnamefont{Japaridze}},
  \bibnamefont{and} \bibinfo{author}{\bibfnamefont{I.~G.}
  \bibnamefont{Kimeridze}}, \bibinfo{journal}{J. Phys. Cond. Mat.}
  \textbf{\bibinfo{volume}{3}}, \bibinfo{pages}{3353} (\bibinfo{year}{1991}).

\bibitem{alfa}
\bibinfo{author}{\bibfnamefont{B.}~\bibnamefont{D\'ora}},
  \bibinfo{author}{\bibfnamefont{K.}~\bibnamefont{Maki}},
  \bibinfo{author}{\bibfnamefont{B.}~\bibnamefont{Korin-Hamzi\'c}},
  \bibinfo{author}{\bibfnamefont{M.}~\bibnamefont{Basleti\'c}},
  \bibinfo{author}{\bibfnamefont{A.}~\bibnamefont{Virosztek}},
  \bibinfo{author}{\bibfnamefont{M.~V.} \bibnamefont{Kartsovnik}},
  \bibnamefont{and}
  \bibinfo{author}{\bibfnamefont{H.}~\bibnamefont{M{\"u}ller}},
  \bibinfo{journal}{Europhys. Lett.} \textbf{\bibinfo{volume}{60}},
  \bibinfo{pages}{737} (\bibinfo{year}{2002}).

\bibitem{prl}
\bibinfo{author}{\bibfnamefont{K.}~\bibnamefont{Maki}},
  \bibinfo{author}{\bibfnamefont{B.}~\bibnamefont{D\'ora}},
  \bibinfo{author}{\bibfnamefont{M.}~\bibnamefont{Kartsovnik}},
  \bibinfo{author}{\bibfnamefont{A.}~\bibnamefont{Virosztek}},
  \bibinfo{author}{\bibfnamefont{B.}~\bibnamefont{Korin-Hamzi\'c}},
  \bibnamefont{and}
  \bibinfo{author}{\bibfnamefont{M.}~\bibnamefont{Basleti\'c}},
  \bibinfo{journal}{Phys. Rev. Lett.} \textbf{\bibinfo{volume}{90}},
  \bibinfo{pages}{256402} (\bibinfo{year}{2003}).

\bibitem{nernst}
\bibinfo{author}{\bibfnamefont{B.}~\bibnamefont{D\'ora}},
  \bibinfo{author}{\bibfnamefont{K.}~\bibnamefont{Maki}},
  \bibinfo{author}{\bibfnamefont{A.}~\bibnamefont{V\'anyolos}},
  \bibnamefont{and}
  \bibinfo{author}{\bibfnamefont{A.}~\bibnamefont{Virosztek}},
  \bibinfo{journal}{Phys. Rev. B} \textbf{\bibinfo{volume}{68}},
  \bibinfo{pages}{241102(R)} (\bibinfo{year}{2003}).

\bibitem{choi}
\bibinfo{author}{\bibfnamefont{E.~S.} \bibnamefont{Choi}},
  \bibinfo{author}{\bibfnamefont{J.~S.} \bibnamefont{Brooks}},
  \bibnamefont{and} \bibinfo{author}{\bibfnamefont{J.~S.}
  \bibnamefont{Qualls}}, \bibinfo{journal}{Phys. Rev. B}
  \textbf{\bibinfo{volume}{65}}, \bibinfo{pages}{205119}
  (\bibinfo{year}{2002}).

\bibitem{dressel}
\bibinfo{author}{\bibfnamefont{M.}~\bibnamefont{Dressel}},
  \bibinfo{author}{\bibfnamefont{G.}~\bibnamefont{Gr{\"u}ner}},
  \bibinfo{author}{\bibfnamefont{J.~P.} \bibnamefont{Pouget}},
  \bibinfo{author}{\bibfnamefont{A.}~\bibnamefont{Breining}}, \bibnamefont{and}
  \bibinfo{author}{\bibfnamefont{D.}~\bibnamefont{Schweitzer}},
  \bibinfo{journal}{J. Phys. I France} \textbf{\bibinfo{volume}{4}},
  \bibinfo{pages}{579} (\bibinfo{year}{1994}).

\bibitem{castroneto}
\bibinfo{author}{\bibfnamefont{A.~H.} \bibnamefont{Castro-Neto}},
  \bibinfo{journal}{Phys. Rev. Lett.} \textbf{\bibinfo{volume}{86}},
  \bibinfo{pages}{4382} (\bibinfo{year}{2001}).

\bibitem{IO}
\bibinfo{author}{\bibfnamefont{H.}~\bibnamefont{Ikeda}} \bibnamefont{and}
  \bibinfo{author}{\bibfnamefont{Y.}~\bibnamefont{Ohashi}},
  \bibinfo{journal}{Phys. Rev. Lett.} \textbf{\bibinfo{volume}{81}},
  \bibinfo{pages}{3723} (\bibinfo{year}{1998}).

\bibitem{bel1}
\bibinfo{author}{\bibfnamefont{R.}~\bibnamefont{Bel}},
  \bibinfo{author}{\bibfnamefont{K.}~\bibnamefont{Behnia}}, \bibnamefont{and}
  \bibinfo{author}{\bibfnamefont{H.}~\bibnamefont{Berger}},
  \bibinfo{journal}{Phys. Rev. Lett.} \textbf{\bibinfo{volume}{91}},
  \bibinfo{pages}{066602} (\bibinfo{year}{2003}).

\bibitem{amitsuka}
\bibinfo{author}{\bibfnamefont{H.}~\bibnamefont{Amitsuka}},
  \bibinfo{author}{\bibfnamefont{M.}~\bibnamefont{Sato}},
  \bibinfo{author}{\bibfnamefont{N.}~\bibnamefont{Metoki}},
  \bibinfo{author}{\bibfnamefont{M.}~\bibnamefont{Yokoyama}},
  \bibinfo{author}{\bibfnamefont{K.}~\bibnamefont{Kuwahara}},
  \bibinfo{author}{\bibfnamefont{T.}~\bibnamefont{Sakakibara}},
  \bibinfo{author}{\bibfnamefont{H.}~\bibnamefont{Morimoto}},
  \bibinfo{author}{\bibfnamefont{S.}~\bibnamefont{Kawarazaki}},
  \bibinfo{author}{\bibfnamefont{Y.}~\bibnamefont{Miyako}}, \bibnamefont{and}
  \bibinfo{author}{\bibfnamefont{J.~A.} \bibnamefont{Mydosh}},
  \bibinfo{journal}{Phys. Rev. Lett.} \textbf{\bibinfo{volume}{83}},
  \bibinfo{pages}{5114} (\bibinfo{year}{1999}).

\bibitem{capan}
\bibinfo{author}{\bibfnamefont{C.}~\bibnamefont{Capan}},
  \bibinfo{author}{\bibfnamefont{K.}~\bibnamefont{Behnia}},
  \bibinfo{author}{\bibfnamefont{J.}~\bibnamefont{Hinderer}},
  \bibinfo{author}{\bibfnamefont{A.~G.~M.} \bibnamefont{Jansen}},
  \bibinfo{author}{\bibfnamefont{W.}~\bibnamefont{Lang}},
  \bibinfo{author}{\bibfnamefont{C.}~\bibnamefont{Marcenat}},
  \bibinfo{author}{\bibfnamefont{C.}~\bibnamefont{Marin}}, \bibnamefont{and}
  \bibinfo{author}{\bibfnamefont{J.}~\bibnamefont{Flouquet}},
  \bibinfo{journal}{Phys. Rev. Lett.} \textbf{\bibinfo{volume}{88}},
  \bibinfo{pages}{056601} (\bibinfo{year}{2002}).

\bibitem{wang1}
\bibinfo{author}{\bibfnamefont{Y.}~\bibnamefont{Wang}},
  \bibinfo{author}{\bibfnamefont{Z.~A.} \bibnamefont{Xu}},
  \bibinfo{author}{\bibfnamefont{T.}~\bibnamefont{Kakeshita}},
  \bibinfo{author}{\bibfnamefont{S.}~\bibnamefont{Uchida}},
  \bibinfo{author}{\bibfnamefont{S.}~\bibnamefont{Ono}},
  \bibinfo{author}{\bibfnamefont{Y.}~\bibnamefont{Ando}}, , \bibnamefont{and}
  \bibinfo{author}{\bibfnamefont{N.~P.} \bibnamefont{Ong}},
  \bibinfo{journal}{Phys. Rev. B} \textbf{\bibinfo{volume}{64}},
  \bibinfo{pages}{224519} (\bibinfo{year}{2001}).

\bibitem{wang2}
\bibinfo{author}{\bibfnamefont{Y.}~\bibnamefont{Wang}},
  \bibinfo{author}{\bibfnamefont{N.~P.} \bibnamefont{Ong}},
  \bibinfo{author}{\bibfnamefont{Z.~A.} \bibnamefont{Xu}},
  \bibinfo{author}{\bibfnamefont{T.}~\bibnamefont{Kakeshita}},
  \bibinfo{author}{\bibfnamefont{S.}~\bibnamefont{Uchida}},
  \bibinfo{author}{\bibfnamefont{D.~A.} \bibnamefont{Bonn}},
  \bibinfo{author}{\bibfnamefont{R.}~\bibnamefont{Liang}}, , \bibnamefont{and}
  \bibinfo{author}{\bibfnamefont{W.~N.} \bibnamefont{Hardy}},
  \bibinfo{journal}{Phys. Rev. Lett.} \textbf{\bibinfo{volume}{88}},
  \bibinfo{pages}{257003} (\bibinfo{year}{2002}).

\bibitem{korea}
\bibinfo{author}{\bibfnamefont{K.}~\bibnamefont{Maki}},
  \bibinfo{author}{\bibfnamefont{B.}~\bibnamefont{D\'ora}},
  \bibinfo{author}{\bibfnamefont{A.}~\bibnamefont{V\'anyolos}},
  \bibnamefont{and}
  \bibinfo{author}{\bibfnamefont{A.}~\bibnamefont{Virosztek}},
  \bibinfo{note}{{C}urr. Appl. Phys. (in press)}.

\bibitem{nambu}
\bibinfo{author}{\bibfnamefont{Y.}~\bibnamefont{Nambu}},
  \bibinfo{journal}{Phys. Rev.} \textbf{\bibinfo{volume}{117}},
  \bibinfo{pages}{648} (\bibinfo{year}{1960}).

\bibitem{impurd-wave}
\bibinfo{author}{\bibfnamefont{Y.}~\bibnamefont{Sun}} \bibnamefont{and}
  \bibinfo{author}{\bibfnamefont{K.}~\bibnamefont{Maki}},
  \bibinfo{journal}{Phys. Rev. B} \textbf{\bibinfo{volume}{51}},
  \bibinfo{pages}{6059} (\bibinfo{year}{1995}).

\bibitem{hotta}
\bibinfo{author}{\bibfnamefont{T.}~\bibnamefont{Hotta}}, \bibinfo{journal}{J.
  Phys. Soc. Jpn.} \textbf{\bibinfo{volume}{62}}, \bibinfo{pages}{274}
  (\bibinfo{year}{1993}).

\bibitem{universal}
\bibinfo{author}{\bibfnamefont{Y.}~\bibnamefont{Sun}} \bibnamefont{and}
  \bibinfo{author}{\bibfnamefont{K.}~\bibnamefont{Maki}},
  \bibinfo{journal}{Europhys. Lett.} \textbf{\bibinfo{volume}{32}},
  \bibinfo{pages}{355} (\bibinfo{year}{1995}).

\bibitem{scatter1}
\bibinfo{author}{\bibfnamefont{B.}~\bibnamefont{D\'ora}},
  \bibinfo{author}{\bibfnamefont{A.}~\bibnamefont{Virosztek}},
  \bibnamefont{and} \bibinfo{author}{\bibfnamefont{K.}~\bibnamefont{Maki}},
  \bibinfo{journal}{Phys. Rev. B} \textbf{\bibinfo{volume}{66}},
  \bibinfo{pages}{115112} (\bibinfo{year}{2002}).

\bibitem{scatter2}
\bibinfo{author}{\bibfnamefont{B.}~\bibnamefont{D\'ora}},
  \bibinfo{author}{\bibfnamefont{A.}~\bibnamefont{Virosztek}},
  \bibnamefont{and} \bibinfo{author}{\bibfnamefont{K.}~\bibnamefont{Maki}},
  \bibinfo{journal}{Phys. Rev. B} \textbf{\bibinfo{volume}{68}},
  \bibinfo{pages}{075104} (\bibinfo{year}{2003}).

\bibitem{yamaji1}
\bibinfo{author}{\bibfnamefont{K.}~\bibnamefont{Yamaji}}, \bibinfo{journal}{J.
  Phys. Soc. Japan} \textbf{\bibinfo{volume}{51}}, \bibinfo{pages}{2787}
  (\bibinfo{year}{1982}).

\bibitem{huang2}
\bibinfo{author}{\bibfnamefont{X.}~\bibnamefont{Huang}} \bibnamefont{and}
  \bibinfo{author}{\bibfnamefont{K.}~\bibnamefont{Maki}},
  \bibinfo{journal}{Phys. Rev. B} \textbf{\bibinfo{volume}{46}},
  \bibinfo{pages}{162} (\bibinfo{year}{1992}).

\bibitem{landau1}
\bibinfo{author}{\bibfnamefont{L.~D.} \bibnamefont{Landau}},
  \bibinfo{journal}{Soviet Phys. JETP} \textbf{\bibinfo{volume}{3}},
  \bibinfo{pages}{920} (\bibinfo{year}{1957}).

\bibitem{landau2}
\bibinfo{author}{\bibfnamefont{L.~D.} \bibnamefont{Landau}},
  \bibinfo{journal}{Soviet Phys. JETP} \textbf{\bibinfo{volume}{5}},
  \bibinfo{pages}{101} (\bibinfo{year}{1957}).

\bibitem{landau3}
\bibinfo{author}{\bibfnamefont{L.~D.} \bibnamefont{Landau}},
  \bibinfo{journal}{Soviet Phys. JETP} \textbf{\bibinfo{volume}{8}},
  \bibinfo{pages}{70} (\bibinfo{year}{1959}).

\bibitem{klasszikus}
\bibinfo{author}{\bibfnamefont{A.~A.} \bibnamefont{Abrikosov}},
  \bibinfo{author}{\bibfnamefont{L.~P.} \bibnamefont{Gor'kov}},
  \bibnamefont{and} \bibinfo{author}{\bibfnamefont{I.~E.}
  \bibnamefont{Dzyaloshinski}}, \emph{\bibinfo{title}{Methods of Quantum Field
  Theory in Statistical Physics}} (\bibinfo{publisher}{Dover Publications},
  \bibinfo{address}{New York}, \bibinfo{year}{1963}).

\bibitem{heisenberg}
\bibinfo{author}{\bibfnamefont{W.}~\bibnamefont{Heisenberg}} \bibnamefont{and}
  \bibinfo{author}{\bibfnamefont{H.}~\bibnamefont{Euler}}, \bibinfo{journal}{Z.
  Phys.} \textbf{\bibinfo{volume}{98}}, \bibinfo{pages}{714}
  (\bibinfo{year}{1936}).

\bibitem{weisskopf}
\bibinfo{author}{\bibnamefont{V.S.Weisskopf}}, \bibinfo{journal}{Kongelige
  Danske Videns. Selskab Mathematisk-fsiske Meddelelser}
  \textbf{\bibinfo{volume}{6}}, \bibinfo{pages}{14} (\bibinfo{year}{1936}).

\bibitem{osada}
\bibinfo{author}{\bibfnamefont{T.}~\bibnamefont{Osada}},
  \bibinfo{author}{\bibfnamefont{A.}~\bibnamefont{Kawasumi}},
  \bibinfo{author}{\bibfnamefont{S.}~\bibnamefont{Kagoshima}},
  \bibinfo{author}{\bibfnamefont{N.}~\bibnamefont{Miura}}, \bibnamefont{and}
  \bibinfo{author}{\bibfnamefont{G.}~\bibnamefont{Saito}},
  \bibinfo{journal}{Phys. Rev. Lett.} \textbf{\bibinfo{volume}{66}},
  \bibinfo{pages}{1525} (\bibinfo{year}{1991}).

\bibitem{naughton}
\bibinfo{author}{\bibfnamefont{M.~J.} \bibnamefont{Naughton}},
  \bibinfo{author}{\bibfnamefont{O.~H.} \bibnamefont{Chung}},
  \bibinfo{author}{\bibfnamefont{M.}~\bibnamefont{Chaparala}},
  \bibinfo{author}{\bibfnamefont{X.}~\bibnamefont{Bu}}, \bibnamefont{and}
  \bibinfo{author}{\bibfnamefont{P.}~\bibnamefont{Coppens}},
  \bibinfo{journal}{Phys. Rev. Lett.} \textbf{\bibinfo{volume}{67}},
  \bibinfo{pages}{3712} (\bibinfo{year}{1991}).

\bibitem{kang1}
\bibinfo{author}{\bibfnamefont{W.}~\bibnamefont{Kang}},
  \bibinfo{author}{\bibfnamefont{S.~T.} \bibnamefont{Hannahs}},
  \bibnamefont{and} \bibinfo{author}{\bibfnamefont{P.~M.}
  \bibnamefont{Chaikin}}, \bibinfo{journal}{Phys. Rev. Lett.}
  \textbf{\bibinfo{volume}{69}}, \bibinfo{pages}{2827} (\bibinfo{year}{1992}).

\bibitem{lee}
\bibinfo{author}{\bibfnamefont{I.~J.} \bibnamefont{Lee}} \bibnamefont{and}
  \bibinfo{author}{\bibfnamefont{M.~J.} \bibnamefont{Naughton}},
  \bibinfo{journal}{Phys. Rev. B} \textbf{\bibinfo{volume}{58}},
  \bibinfo{pages}{R13343} (\bibinfo{year}{1998}).

\bibitem{kang2}
\bibinfo{author}{\bibfnamefont{W.}~\bibnamefont{Kang}},
  \bibinfo{author}{\bibfnamefont{H.-Y.} \bibnamefont{Kang}},
  \bibinfo{author}{\bibfnamefont{Y.~J.} \bibnamefont{Jo}}, \bibnamefont{and}
  \bibinfo{author}{\bibfnamefont{S.}~\bibnamefont{Uji}},
  \bibinfo{journal}{Synth. Met.} \textbf{\bibinfo{volume}{133}},
  \bibinfo{pages}{15} (\bibinfo{year}{2003}).

\bibitem{lebed1}
\bibinfo{author}{\bibfnamefont{A.~G.} \bibnamefont{Lebed}},
  \bibinfo{journal}{JETP Lett.} \textbf{\bibinfo{volume}{43}},
  \bibinfo{pages}{174} (\bibinfo{year}{1986}).

\bibitem{lebed2}
\bibinfo{author}{\bibfnamefont{A.~G.} \bibnamefont{Lebed}} \bibnamefont{and}
  \bibinfo{author}{\bibfnamefont{P.}~\bibnamefont{Bak}},
  \bibinfo{journal}{Phys. Rev. Lett.} \textbf{\bibinfo{volume}{63}},
  \bibinfo{pages}{1315} (\bibinfo{year}{1989}).

\bibitem{valami}
\bibinfo{author}{\bibfnamefont{B.}~\bibnamefont{D\'ora}},
  \bibinfo{author}{\bibfnamefont{K.}~\bibnamefont{Maki}}, \bibnamefont{and}
  \bibinfo{author}{\bibfnamefont{A.}~\bibnamefont{Virosztek}},
  \bibinfo{note}{in preparation}.

\bibitem{basleticprb}
\bibinfo{author}{\bibfnamefont{M.}~\bibnamefont{Basleti\'c}},
  \bibinfo{author}{\bibfnamefont{B.}~\bibnamefont{Korin-Hamzi\'c}},
  \bibnamefont{and} \bibinfo{author}{\bibfnamefont{K.}~\bibnamefont{Maki}},
  \bibinfo{journal}{Phys. Rev. B} \textbf{\bibinfo{volume}{65}},
  \bibinfo{pages}{235117} (\bibinfo{year}{2002}).

\bibitem{petrovic}
\bibinfo{author}{\bibfnamefont{M.}~\bibnamefont{Nicklas}},
  \bibinfo{author}{\bibfnamefont{R.}~\bibnamefont{Borth}},
  \bibinfo{author}{\bibfnamefont{E.}~\bibnamefont{Lengyel}},
  \bibinfo{author}{\bibfnamefont{P.~G.} \bibnamefont{Pagliuso}},
  \bibinfo{author}{\bibfnamefont{J.~L.} \bibnamefont{Sarrao}},
  \bibinfo{author}{\bibfnamefont{V.~A.} \bibnamefont{Sidorov}},
  \bibinfo{author}{\bibfnamefont{G.}~\bibnamefont{Sparn}},
  \bibinfo{author}{\bibfnamefont{F.}~\bibnamefont{Steglich}}, \bibnamefont{and}
  \bibinfo{author}{\bibfnamefont{J.~D.} \bibnamefont{Thompson}},
  \bibinfo{journal}{J. Phys. Cond. Matt.} \textbf{\bibinfo{volume}{13}},
  \bibinfo{pages}{L905} (\bibinfo{year}{2001}).

\bibitem{izawa}
\bibinfo{author}{\bibfnamefont{K.}~\bibnamefont{Izawa}},
  \bibinfo{author}{\bibfnamefont{H.}~\bibnamefont{Yamaguchi}},
  \bibinfo{author}{\bibfnamefont{Y.}~\bibnamefont{Matsuda}},
  \bibinfo{author}{\bibfnamefont{H.}~\bibnamefont{Shishido}},
  \bibinfo{author}{\bibfnamefont{R.}~\bibnamefont{Settai}}, \bibnamefont{and}
  \bibinfo{author}{\bibfnamefont{Y.}~\bibnamefont{Onuki}},
  \bibinfo{journal}{Phys. Rev. Lett.} \textbf{\bibinfo{volume}{87}},
  \bibinfo{pages}{057002} (\bibinfo{year}{2001}).

\bibitem{shishida}
\bibinfo{author}{\bibfnamefont{I.}~\bibnamefont{Shishido}},
  \bibinfo{author}{\bibfnamefont{R.}~\bibnamefont{Settai}},
  \bibinfo{author}{\bibfnamefont{D.}~\bibnamefont{Aoki}},
  \bibinfo{author}{\bibfnamefont{S.}~\bibnamefont{Ikeda}},
  \bibinfo{author}{\bibfnamefont{H.}~\bibnamefont{Nakawaki}},
  \bibinfo{author}{\bibfnamefont{N.}~\bibnamefont{Nakamura}},
  \bibinfo{author}{\bibfnamefont{T.}~\bibnamefont{Iizuka}},
  \bibinfo{author}{\bibfnamefont{Y.}~\bibnamefont{Inada}},
  \bibinfo{author}{\bibfnamefont{K.}~\bibnamefont{Sugiyama}},
  \bibinfo{author}{\bibfnamefont{T.}~\bibnamefont{Takeuchi}},
  \bibinfo{author}{\bibfnamefont{K.}~\bibnamefont{Kindo}},
  \bibinfo{author}{\bibfnamefont{T.~C.} \bibnamefont{Kobayashi}},
  \emph{et~al.}, \bibinfo{journal}{J. Phys. Soc. Jpn.}
  \textbf{\bibinfo{volume}{71}}, \bibinfo{pages}{162} (\bibinfo{year}{2002}).

\bibitem{sidorov}
\bibinfo{author}{\bibfnamefont{V.~A.} \bibnamefont{Sidorov}},
  \bibinfo{author}{\bibfnamefont{M.}~\bibnamefont{Nicklas}},
  \bibinfo{author}{\bibfnamefont{P.~G.} \bibnamefont{Pagliuso}},
  \bibinfo{author}{\bibfnamefont{J.~L.} \bibnamefont{Sarrao}},
  \bibinfo{author}{\bibfnamefont{Y.}~\bibnamefont{Bang}},
  \bibinfo{author}{\bibfnamefont{A.~V.} \bibnamefont{Balatsky}},
  \bibnamefont{and} \bibinfo{author}{\bibfnamefont{J.~D.}
  \bibnamefont{Thompson}}, \bibinfo{journal}{Phys. Rev. Lett.}
  \textbf{\bibinfo{volume}{89}}, \bibinfo{pages}{157004}
  (\bibinfo{year}{2003}).

\bibitem{bel2}
\bibinfo{author}{\bibfnamefont{R.}~\bibnamefont{Bel}},
  \bibinfo{author}{\bibfnamefont{K.}~\bibnamefont{Behnia}},
  \bibinfo{author}{\bibfnamefont{Y.}~\bibnamefont{Nakajima}},
  \bibinfo{author}{\bibfnamefont{K.}~\bibnamefont{Izawa}},
  \bibinfo{author}{\bibfnamefont{Y.}~\bibnamefont{Matsuda}},
  \bibinfo{author}{\bibfnamefont{H.}~\bibnamefont{Shishido}},
  \bibinfo{author}{\bibfnamefont{R.}~\bibnamefont{Settai}}, \bibnamefont{and}
  \bibinfo{author}{\bibfnamefont{Y.}~\bibnamefont{Onuki}},
  \bibinfo{note}{cond-mat/0311473}.

\bibitem{salerno}
\bibinfo{author}{\bibfnamefont{K.}~\bibnamefont{Maki}},
  \bibinfo{author}{\bibfnamefont{B.}~\bibnamefont{D\'ora}}, \bibnamefont{and}
  \bibinfo{author}{\bibfnamefont{A.}~\bibnamefont{Virosztek}}
  (\bibinfo{publisher}{AIP Conference Proceedings 695},
  \bibinfo{address}{Melville}, \bibinfo{year}{2003}), p.~\bibinfo{pages}{10}.

\bibitem{kanoda}
\bibinfo{author}{\bibfnamefont{K.}~\bibnamefont{Kanoda}},
  \bibinfo{journal}{Physica C} \textbf{\bibinfo{volume}{282-287}},
  \bibinfo{pages}{299} (\bibinfo{year}{1997}).

\bibitem{izawa2}
\bibinfo{author}{\bibfnamefont{K.}~\bibnamefont{Izawa}},
  \bibinfo{author}{\bibfnamefont{H.}~\bibnamefont{Yamaguchi}},
  \bibinfo{author}{\bibfnamefont{T.}~\bibnamefont{Sasaki}}, \bibnamefont{and}
  \bibinfo{author}{\bibfnamefont{Y.}~\bibnamefont{Matsuda}},
  \bibinfo{journal}{Phys. Rev. Lett.} \textbf{\bibinfo{volume}{88}},
  \bibinfo{pages}{027002} (\bibinfo{year}{2002}).

\bibitem{muller}
\bibinfo{author}{\bibfnamefont{J.}~\bibnamefont{M{\"u}ller}},
  \bibinfo{author}{\bibfnamefont{M.}~\bibnamefont{Lang}},
  \bibinfo{author}{\bibfnamefont{F.}~\bibnamefont{Steglich}},
  \bibinfo{author}{\bibfnamefont{J.~A.} \bibnamefont{Schlueter}},
  \bibinfo{author}{\bibfnamefont{A.~M.} \bibnamefont{Kini}}, \bibnamefont{and}
  \bibinfo{author}{\bibnamefont{T.Sasaki}}, \bibinfo{journal}{Phys. Rev. B}
  \textbf{\bibinfo{volume}{65}}, \bibinfo{pages}{144521}
  (\bibinfo{year}{2002}).

\bibitem{pinteric}
\bibinfo{author}{\bibfnamefont{M.}~\bibnamefont{Pinteri\'c}},
  \bibinfo{author}{\bibfnamefont{S.}~\bibnamefont{Tomi\'c}},
  \bibinfo{author}{\bibfnamefont{M.}~\bibnamefont{Prester}},
  \bibinfo{author}{\bibfnamefont{D.}~\bibnamefont{Dorba\'c}}, \bibnamefont{and}
  \bibinfo{author}{\bibnamefont{K.Maki}}, \bibinfo{journal}{Phys. Rev. B}
  \textbf{\bibinfo{volume}{66}}, \bibinfo{pages}{174521}
  (\bibinfo{year}{2002}).

\end{thebibliography}

\end{document}